\documentclass[aps,prc,preprint,superscriptaddress,groupedaddress,showpacs,floatfix,amsmath,amssymb,]{revtex4}
\usepackage{graphicx}
\usepackage{subfigure}
\usepackage{esint}

\begin{document}

\title{Fermion propagators in space-time}

\author{M.~B.~Barbaro}
\affiliation{Dipartimento di Fisica Teorica, Universit\`{a} di Torino and INFN,\\
Sezione di Torino, Via P. Giuria 1, 10125 Torino, Italy}

\author{D.~Berardo}
\affiliation{Dipartimento di Fisica Teorica, Universit\`{a} di Torino and INFN,\\
Sezione di Torino, Via P. Giuria 1, 10125 Torino, Italy}

\author{R.~Cenni}
\affiliation{INFN, Sezione di Genova, Via Dodecaneso 33, I-16146 Genova, Italy}

\author{T.~W.~Donnelly}
\affiliation{Center for Theoretical Physics, Laboratory for Nuclear Science and Department of Physics, Massachusetts Institute of
Technology, Cambridge, Massachusetts 02139, USA}

\author{A.~Molinari}
\affiliation{Dipartimento di Fisica Teorica, Universit\`{a} di Torino and INFN,\\
Sezione di Torino, Via P. Giuria 1, 10125 Torino, Italy}

\begin{abstract}

The one- and the two-particle propagators for an infinite
non-interacting Fermi system are studied as functions of space-time
coordinates. Their behaviour at the origin and in the asymptotic
region is discussed, as is their scaling in the Fermi momentum. Both
propagators are shown to have a divergence at equal times. The
impact of the interaction among the fermions on their momentum
distribution, on their pair correlation function and, hence, on the
Coulomb sum rule is explored using a phenomenological model. Finally
the problem of how the confinement is reflected in the momentum
distribution of the system's constituents is briefly addressed.

\end{abstract}

\pacs{24.10.-i, 24.10.Cn, 25.30.-c} \maketitle

\section{Introduction}
\label{sec:intro}

In this study we derive expressions in space-time for the one- and
two-fermion propagators, considering the simple case of a
non-relativistic, non-interacting infinite Fermi system where these
quantities in energy-momentum space are well-known. To obtain the
corresponding space-time results is not an entirely trivial job and
we were unable to find any detailed discussion of these quantities
in the literature. The motivation for the present study comes
principally from the desire to investigate the roles played by
correlations, in particular short-range correlations, in nuclear
matter and finite nuclei, although our study applies as well to
other fields of many-body physics. From this perspective the present
work should be viewed as a first step in the direction of treating
more complex systems: we present arguments later that the
correlations among the constituents of a Fermi~\cite{corr} system
are best appreciated in space-time, especially so when compared with
the non-dynamically-correlated situation. We shall illustrate this
point using a model for the dynamical correlations which modifies
the non-interacting step-function momentum distribution at the Fermi
surface.

Two further items are addressed because of their significance for a
non-interacting system. The first relates to the propagators' second
kind scaling property (independence of the Fermi
momentum)~\cite{DS99l,DS99,BCD+98,Amaro2002}: we prove that,
providing an appropriate rescaling of space and time is performed,
both the one-fermion and two-fermion propagators do scale in $k_F$,
the latter except at equal times. Indeed one finds that when the
quantum field theory (QFT) is applied to a finite-density many-body
system a divergence occurs if the propagators are evaluated at equal
times. One also finds that both the one- and two-particle
propagators become purely imaginary at equal times.

The second item concerns the problem of the impact of confinement on
the non-interacting fermion momentum distribution (and hence on the
propagators).

The present paper is organized as follows. In Sec.~\ref{sec:onebody}
we deal with the one-body propagator $G^0(x,x')$. We observe that,
while the hole propagator is well-defined, at equal times the
particle propagator is not. We derive an analytic expression for the
space-time hole propagator and discuss its asymptotic behaviour. We
find that it vanishes as a power law at large $|\vec{x}-\vec{x}'|$
and $t-t'$ as a consequence of the cut the function $G^0({\vec
k},\omega)$ displays in the complex $\omega$-plane just above the
real axis for $0\le\omega\le\omega_F$, where $\omega_F=k_F^2/2m$,
with $k_F$ the Fermi momentum and $m$ the fermion (nucleon) mass
(Pailey-Wiener theorem~\cite{DeAlfaro}). We also prove that
$G^0(x,x')$ divided by the density scales in $k_F$. In
Sec.~\ref{sec:CSR} we proceed to study the two-particle propagator,
specifically the density-density correlation function, starting by
focusing on the Coulomb Sum Rule (CSR)~\cite{Amore:1996kg}. To
illustrate how such topics can be addressed in space-time, we derive
the well-known expression for the CSR using the pair correlation
function (which arises in the present case simply from the Pauli
principle) as input and performing the integration in the complex
coordinate space. In Sec.~\ref{sec:DDCF} we obtain the space-time
expression for the density-density correlation function, usually
referred to as $\Pi^0(x, x')$. We do this partly in terms of the
error function and partly in terms of a function for which we keep
the integral representation, although it may be expressed in terms
of the Meijer $G$-functions~\cite{Abr}. We analyze the asymptotic
behaviour of $\Pi^0(x, x')$, compare it with that of $G^0(x,x')$ and
discuss its scaling behaviour in $k_F$. Next we show that the
imaginary part of the density-density correlation function (a branch
of the two-particle propagator), unlike the CSR, cannot be expressed
directly through the Pauli pair correlation function, its only
ingredient being the momentum distribution. Furthermore we show that
the QFT shortcoming previously found in the case of $G^0(x,x')$ also
affects $\Pi^0(x, x')$. In Sec.~\ref{sec:modif}, we introduce a
phenomenological momentum distribution, which we employ for
computing the CSR and discuss the significance of the difference
between the latter and the CSR of the free Fermi gas. In the
Conclusions (Sec.~\ref{sec:concl}) we briefly address the problem of
how the confinement of our system affects the momentum distribution
of its constituents,  summarize our findings and outline a few
further important issues we intend to address in future work.

\section{The one-body propagator}
\label{sec:onebody}

In this section we deal with the four-dimensional Fourier transform
of the well-known one-body fermion propagator $G^0({\vec k},\omega)$ in an
infinite, homogenoeus, non-interacting system. That is, we compute
\begin{equation}
\label{eq:G01} G^0(x,x')=\int\frac{d\vec
k}{(2\pi)^3}e^{i\vec{k}\cdot(\vec{x}-\vec{x}')}
\int_{-\infty}^{+\infty}
\frac{d\omega}{2\pi}e^{-i\omega(t-t')}\bigg\{\frac{\theta(k-k_F)}{\omega-\omega_k+i\eta}
+\frac{\theta(k_F-k)}{\omega-\omega_k-i\eta}\bigg\} ~,
\end{equation}
where here for simplicity spin indices are suppressed. The frequency
integration in Eq.~\eqref{eq:G01} is easily performed in the complex
$\omega$-plane and one can recognize forward (particle) and backward
(hole) propagation. In Eq.~\eqref{eq:G01} the angular integrations
are also immediate, and one is left with the expression
\begin{equation}
\label{eq:G02}
G^0(x,x')=\frac{i}{2\pi^2 r}\bigg\{\theta(t'-t)\int_{0}^{k_F}dkke^{-i\omega_k(t-t')}\sin(kr)
-\theta(t-t')\int_{k_F}^{\infty}dkke^{-i\omega_k(t-t')}\sin(kr)\bigg\} ,
\end{equation}
where $r\equiv |\vec{x}-\vec{x}'|$ and $\omega_k=k^2/(2m)$. The
integral
\begin{equation}
\label{eq:int}
\int_{0}^{\infty}dkke^{-i\omega_k(t-t')}\sin(kr)=\frac{\sqrt{\pi}}{4}
\frac{r e^{i\frac{m r^{2}}{2(t-t')}}}{\sqrt{(\frac{i}{2m}(t-t'))^3}}
\end{equation}
diverges when computed at equal times $t=t'$ for any value of $r$,
and thus at equal times the particle propagator in the present
framework is ill-defined and cannot be computed unless regularized.
For instance, this might be achieved in the context of the Wightman
formulation of QFT (see ~\cite{Wightman}) which replaces  the field
function with a distribution.

We do not dwell here on the problem of the regularization of the
particle propagator and instead limit our attention to the
computation of the hole propagator. This can be done analytically
and yields ($\alpha$ and $\beta$ are the spin indices which we
temporarily reintroduce)
\begin{eqnarray}
\label{eq:G04} &&G_{\alpha \beta}^0(x,x')= \delta_{\alpha \beta}
i\theta(t'-t) \frac{n_0}{2}\frac{3}{k_F r}e^{i\Delta_t}
\bigg\{\frac{(k_F r)^2}{(k_F r)^2+2i\Delta_t}\bigg[j_1(k_F r)+
\frac{\cos(k_F r)}{k_F r}+\frac{\sin(k_F r)}{2i\Delta_t}\bigg]
\nonumber\\&& + \frac{\sqrt{\pi}}{2} \frac{k_F r}{(2\Delta_t)^{3/2}}
\; \frac{1-i}{2} e^{- i \big [ \big ( \frac{k_F r}{ 2 \sqrt{
\Delta_t}} \big )^2 + \Delta_t \big ] } \nonumber\\&&\times
\bigg[erf\bigg(\frac{1-i}{2}\bigg(\frac{k_F
r}{\sqrt{2\Delta_t}}-\sqrt{2\Delta_t}\bigg)\bigg)-
erf\bigg(\frac{1-i}{2}\bigg(\frac{k_F
r}{\sqrt{2\Delta_t}}+\sqrt{2\Delta_t}\bigg)\bigg)\bigg] \bigg\}\ ,
\end{eqnarray}
where $n_0=k_F^3/3\pi^2$ is the system's density for spin 1/2
particles  (the case we are considering). Furthermore,
\begin{equation}
\label{eq:Deltat} \Delta_t = \frac{k_F^2}{2m}(t'-t)
\end{equation}
is the time difference expressed in inverse Fermi frequency units
(the natural choice in the present context), $j_1(k_F r)$ is the
spherical Bessel function of order one and the standard definition
\begin{equation}
\label{eq:erf} erf(z) = \frac{2}{\sqrt{\pi}} \int_0^z e^{-t^2} dt
\end{equation}
for the error function is employed. Note that, but for the overall
factor  $n_0$, the Green function $G^0(x,x')$ scales in the system's
density, {\it
 i.e.,} it loses all explicit $k_F$ dependence which
then enters only in defining the scale of the space at any value of
$\Delta_t$.

>From the asymptotic expansion of the $erf(z)$ (see Appendix
\ref{sec:app_prop})  it follows that the equal-time limit of
$G^0(x,x')$ reads
\begin{equation}
\label{eq:G0equalt} G^0({\vec x}t,{\vec {x}'}t^+) = i \frac{n_0}{2} \frac{3}{k_F r} j_1(k_F r)
\end{equation}
which, when $r\to 0$ and the spin trace is taken, reduces to the
system's density, as it should. Notably Eq.~\eqref{eq:G0equalt} also
holds for finite $\Delta_t$ and  large $r$, as is shown in Appendix
\ref{sec:app_prop} where it is proven that under these conditions
the contribution stemming from the two error functions exactly
cancels the one arising from the second and third terms inside the
square brackets in Eq.~\eqref{eq:G04}. Thus, for fixed $\Delta_t$
the propagator in Eq.~\eqref{eq:G04} asymptotically behaves as
$\cos(k_F r)/(k_F r)^2$, reflecting the square root singularity
$k=\sqrt{2m\omega}-i\epsilon$ of $G^0({\vec k},\omega)$ in the
complex $k$-plane. Indeed  as illustrated in \cite{DeAlfaro}, which
deals with the theory of potential scattering, the structure of the
Fourier transform basically implies that the transform of a
singularity {\it is} an asymptotic behaviour. Analogously $G^0({\vec
k},\omega)$ displays a $[\Delta_t]^{-1}$ behaviour at large times
for fixed $r$, since in the complex $\omega$-plane the propagator
has a simple pole for fixed ${\vec k}$.

\begin{figure}[htbp]
\centering
\includegraphics[height=5cm]{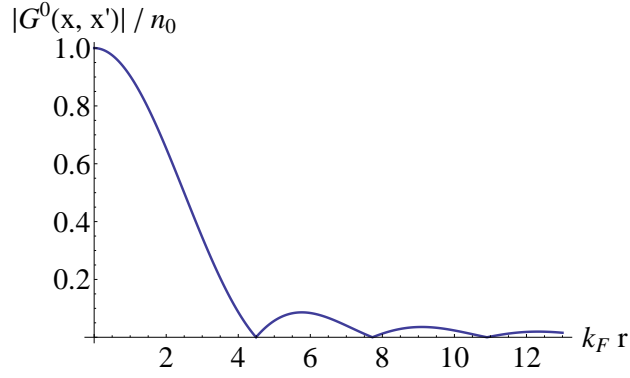}
\caption{\label{fig:g00} The equal-time hole propagator
normalized to the density $n_0$ as given in
Eq.~\eqref{eq:G0equalt}  versus $k_F r$ ($r$ is the modulus of the
relative distance).}
\end{figure}

\begin{figure}[htbp]
\centering%
\subfigure[\label{fig:g05re}]%
{\includegraphics[height=4cm]{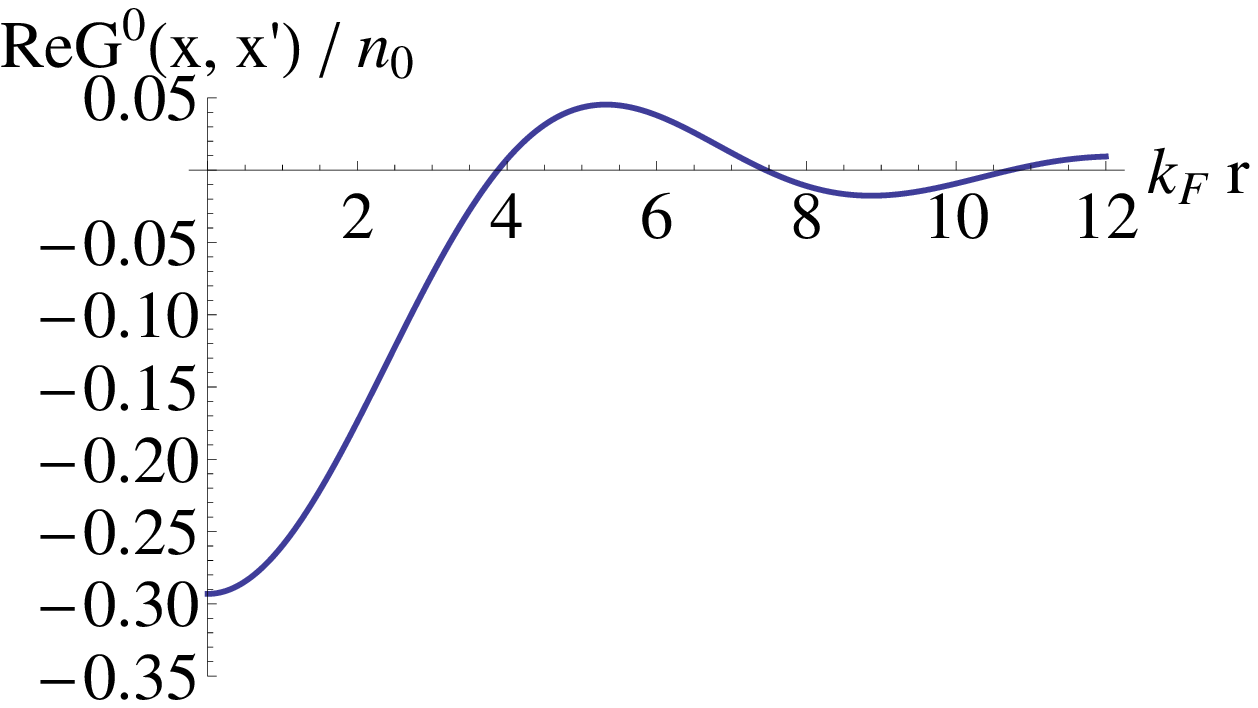}}\qquad\qquad
\subfigure[\label{fig:g05im}]%
{\includegraphics[height=4cm]{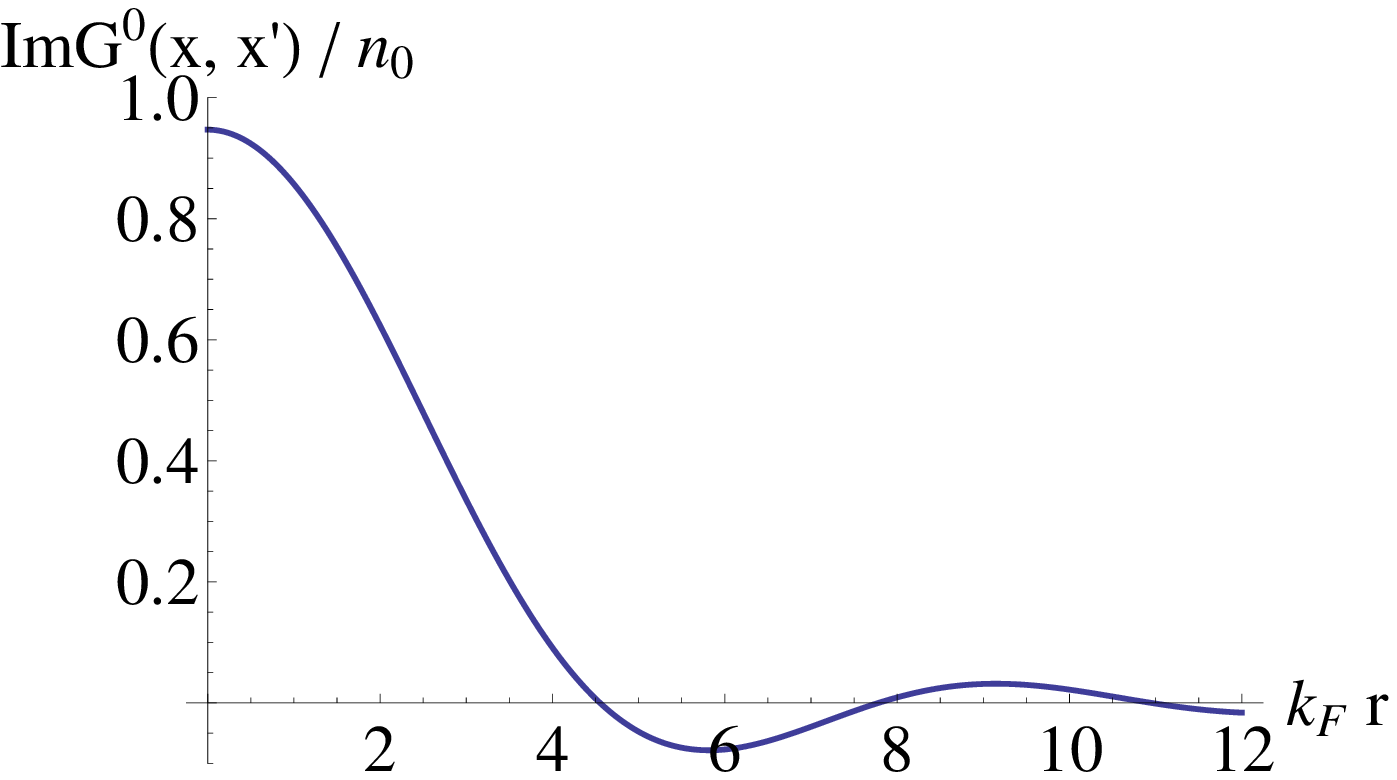}}\qquad\qquad
\subfigure[\label{fig:g05abs}]%
{\includegraphics[height=4cm]{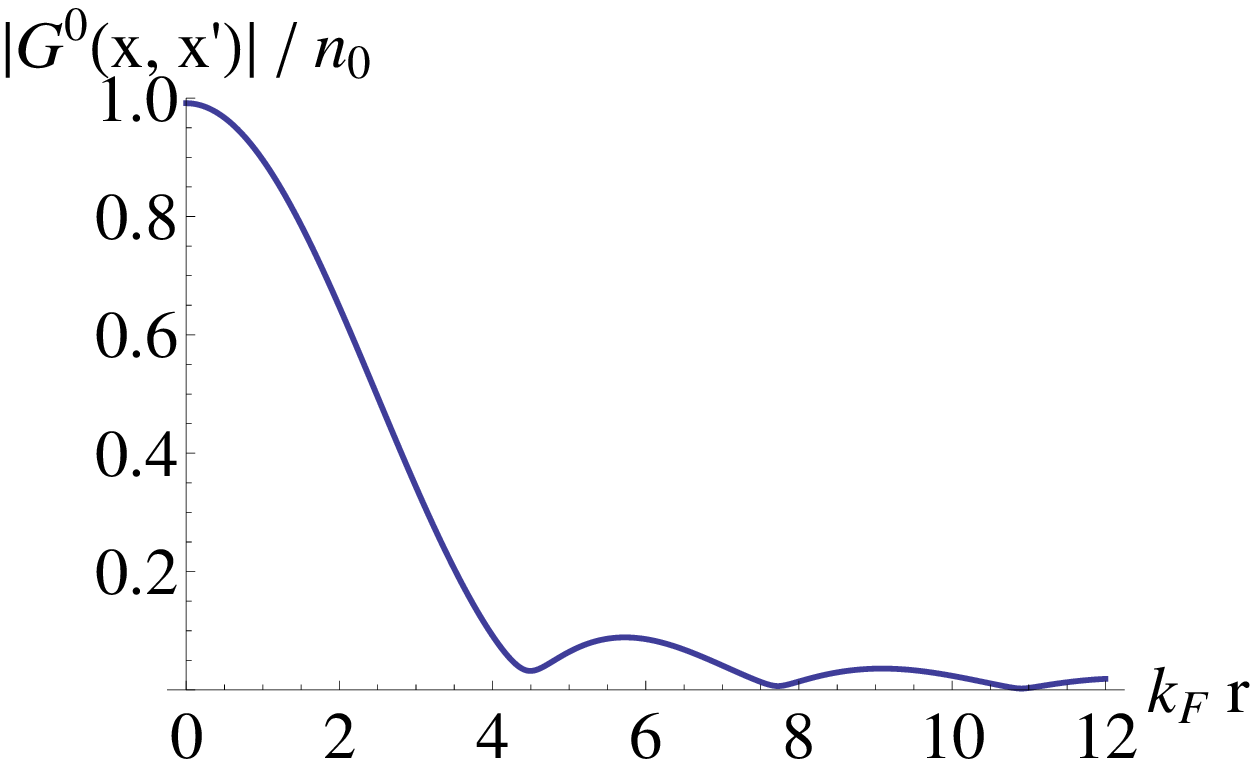}} \caption{ \label{fig:g05}
Real part, imaginary part and modulus of $G^0(x,x')$
normalized to the density $n_0$ for $\Delta_t =0.5 $~.}
\end{figure}

\begin{figure}[htbp]
\centering%
\subfigure[\label{fig:g2re}]%
{\includegraphics[height=4cm]{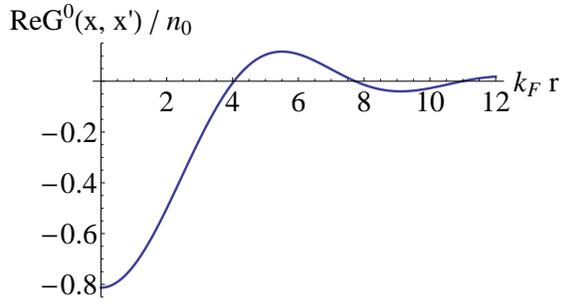}}\qquad\qquad
\subfigure[\label{fig:g2im}]%
{\includegraphics[height=4cm]{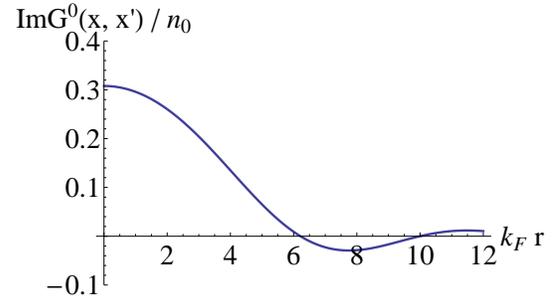}}\qquad\qquad
\subfigure[\label{fig:g2abs}]%
{\includegraphics[height=4cm]{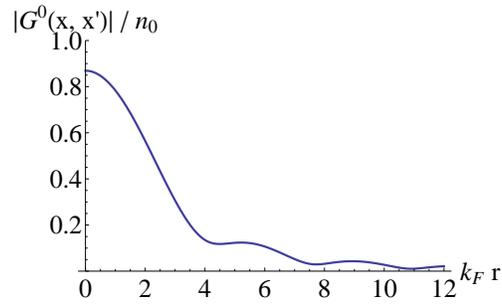}} \caption{ \label{fig:g1}
Same as in Fig.~\ref{fig:g05}, but for $\Delta_t = 2$~.}
\end{figure}

The above findings are borne out by the results shown in
Figs.~\ref{fig:g00}, ~\ref{fig:g05} and ~\ref{fig:g1},  where one
sees the behaviour of the modulus of $G^0(x,x')$ versus $k_F r$ for
a few values of $\Delta_t$. Also shown in Figs.~\ref{fig:g05} and
~\ref{fig:g1} are the real and imaginary parts of $G^0(x,x')$ for
two finite values of $\Delta_t$. As is well-known in $({\vec
k},\omega)$ space, these are connected through a dispersion
relation.

Importantly for $\Delta_t = 0$ the  hole propagator becomes purely
imaginary and  displays an oscillatory behaviour versus $k_F r$.
This accounts for the zeros in its modulus shown in
Fig.~\ref{fig:g00}. From the point of view of the Heisenberg
principle $\Delta_t = 0$ corresponds to the maximum energy (in fact
momentum) uncertainty for the propagating particle. Thus the wave
function of the latter corresponds to a large superposition of plane
waves yielding the striking zeros seen in Fig.~\ref{fig:g00}. For
$\Delta_t$ small, but not vanishing (Fig.~\ref{fig:g05}), the
propagator starts to develop a real part. This also oscillates, but
its zeros differ from those of the imaginary part. Accordingly, now
the particle can be found everywhere in space, but of course with a
small probability when in the vicinity of the zeros appearing in
Fig.~\ref{fig:g00}. As $\Delta_t$ grows (Fig.~\ref{fig:g1}) so does
the real part of $G^0(x,x')$ and the behaviour of the modulus of the
propagator becomes smoother and smoother, until for $\Delta_t$ very
large it becomes constant: indeed now, in accord with the Heisenberg
principle, the wave function of the particle becomes a plane wave.

\section{The Coulomb sum rule}
\label{sec:CSR}

To pave the way to the more general treatment of the two-particle
propagator  (of which the density-density correlation function is a
particular  branch) to be discussed in the next section, we once
more derive the well-known Coulomb sum rule, although this time
through a somewhat different technique, namely an integration in
coordinate space. For this purpose we recall (see ~\cite{FW})
that, using standard quantum mechanics, the CSR can be cast in the
form
\begin{equation}
\label{eq:CSR1} S(\vec q)= <\Psi_0|\tilde{\rho}^\dagger(-\vec
q)\tilde{\rho}(-\vec q)|\Psi_0>\ ,
\end{equation}
where $|\Psi_0>$ is the system's ground state and $\tilde\rho(-\vec
q)$ is the density  deviation operator defined as
\begin{equation}
\label{eq:rho}
\tilde{\rho}(-\vec q)= \hat{\rho}(-\vec q)- <\Psi_0|\hat{\rho}(- \vec q)|\Psi_0> ~.
\end{equation}
It is then a straightforward matter to obtain from Eq.~\eqref{eq:CSR1} the formula
\begin{equation}
\label{eq:CSR2}
S(\vec q) = Z + \int d \vec {x} d \vec {y} e^{-i \vec{q} \cdot (\vec{x}-\vec{y})}
<\Psi_0|\hat{\Psi}_{\alpha}^{\dagger}(\vec x) \hat{\Psi}_{\beta}^{\dagger}(\vec y) \hat{\Psi}_{\beta}(\vec y)
\hat{\Psi}_{\alpha} (\vec x)|\Psi_0>-\rho_0^2V\delta(\vec q)
\end{equation}
which expresses the CSR in terms of the fermion fields in the
Schroedinger picture  (spin indices have been reintroduced) always
sticking to the model of a homogeneous system enclosed in a large
volume $V$ having $Z=A/2$ charges, with $A$ being the total number
of spin $1/2$ particles.
In the second term on the right-hand side of Eq.~\eqref{eq:CSR2} one
recognizes the equal-time two-particle propagator. This can be
expressed in terms of the correlations among the particles in the
system and indeed, in our simple case, it reads
\begin{equation}
\label{eq:corrf} <\Psi_0|\hat{\Psi}_{\alpha}^{\dagger}(\vec
x)\hat{\Psi}_{\beta}^{\dagger}(\vec y)  \hat{\Psi}_{\beta}(\vec y)
\hat{\Psi}_{\alpha} (\vec
x)|\Psi_0>=n_0^2\bigg[1-\frac{1}{2}\bigg(\frac{3 j_{1}(k_F r)}{k_F
r}\bigg)^2\bigg] = n_0^2 \Big [ 1 - \frac{1}{2} g^2(k_Fr) \Big ]
\end{equation}
with $r\equiv |\vec{x} -\vec{y}|$. The function $\frac{1}{2} g^2(k_F
r)$ is usually referred to as the  Pauli pair correlation function.
In the above the direct and exchange contributions clearly appear.
Moreover it  should be kept in mind that basic in deriving
Eq.~\eqref{eq:corrf} have been the $\theta$-functions entering in
$G^0({\vec k},\omega)$. Our aim here is to show that even small
variations of Eq.~\eqref{eq:corrf} at short distances (and hence of
the $\theta$-functions in momentum space) can produce quite dramatic
changes in the CSR (see Sec.~\ref{sec:modif}).

We proceed by inserting Eq.~\eqref{eq:corrf} into
Eq.~\eqref{eq:CSR2}, getting  for the latter
\begin{eqnarray}
\label{eq:CSR3} S(\vec q)&=& Z+ n_0^2 V \delta(\vec q)-n_0^2 V \int
d\vec{r} e^{-i \vec{q}  \cdot \vec{r}} \frac{1}{2}
\bigg(\frac{3j_{1}(k_F r)}{k_F
r}\bigg)^2 - n_0^2 V \delta(\vec q)\nonumber\\
&=& Z \bigg[ 1- \frac{6}{\pi}\int_{0}^{\infty}dz j_{1}^{2}(z) j_{0}
\bigg(\frac{q}{k_F}z\bigg)\bigg]\ ,
\end{eqnarray}
where the elementary angular integrations have been performed and
the term  arising from the direct piece of Eq.~\eqref{eq:corrf} is
seen to drop out.

The integral in Eq.~\eqref{eq:CSR3} can be computed in the complex $z$-plane using standard
techniques  (see Appendix \ref{sec:app_integral} for details),
yielding for the Coulomb sum rule the familiar non-relativistic
expression~\cite{FW}
\begin{eqnarray}
\label{eq:CSR4}
S(\vec q) &=&
\begin{cases}
Z \Big ( \frac{3}{4} \frac{q}{k_F} - \frac{1}{2} \Big(\frac{q}{2 k_F}  \Big)^3 \Big ) & \text {if $q < 2 k_F$} \\
Z  & \text {if $q \geq 2 k_F$}   ~.
\end{cases}
\end{eqnarray}
In connection with the above derivation which is valid for a perfect
Fermi gas,  it should be pointed out that  the result in
Eq.~\eqref{eq:CSR4} stems from the exact cancellation of two
contributions, as shown in Appendix B. We shall then prove in
Section \ref{sec:modif} that, as anticipated above, even a minor
modification induced by interactions among the system's constituents
of the $\theta$-functions entering in $G^0({\vec k},\omega)$ is
sufficient to disrupt the cancellation in Eq.~\eqref{eq:I8}: hence
such a modification induces a sizable change of the CSR for large
$q$. Since, as will be shown in Sec.~\ref{sec:modif}, modifying the
$\theta$-function around the value $k=k_F$ actually corresponds to
modifying the pair distribution in Eq.~\eqref{eq:corrf} for small
distances, this outcome is what one should expect and it offers a
nice example of how the Fourier transform works.

\section{The density-density correlation function}
\label{sec:DDCF}

In this section we analytically compute and explore in space-time
the branch of the two-particle propagator usually referred to as the
density-density correlation function or polarization propagator
$\Pi(x,y)$. The expression for the latter is again well-known in
energy-momentum space  for a non-interacting system, as is the fact
that its imaginary part provides the inelastic scattering cross
section for many types of probes of a many-body system.

The definition of $\Pi(x,0)$ is the following
\begin{equation}
\label{eq:Pi1}
\Pi(x,0)=-i<\Psi_0|T(\tilde{\rho}_{H}(x)\tilde{\rho}_{H}(0))|\Psi_0>\
,
\end{equation}
where the density deviation operators are in the Heisenberg picture,
unlike  the case of the CSR where they were taken in the
Schroedinger picture, {\it i.e.,} where no T-product was introduced.
To explore this point further we recast Eq.~\eqref{eq:Pi1} in the
form
\begin{eqnarray}
\label{eq:Pi2}
\Pi(x,0)=-i&\{&<\Psi_0|T(\hat{\Psi}_{\alpha}^{\dagger}(x)\hat{\Psi}_{\alpha}(x)\hat{\Psi}_{\beta}^{\dagger}(0)
\hat{\Psi}_{\beta}(0) )|\Psi_0> \nonumber\\
&-&<\Psi_0|\hat{\Psi}_{\alpha}^{\dagger}(x)\hat{\Psi}_{\alpha}(x)|\Psi_0><\Psi_0|
\hat{\Psi}_{\beta}^{\dagger}(0)\hat{\Psi}_{\beta}(0)|\Psi_0>\} ~.
\end{eqnarray}
A comparison with Eq.~\eqref{eq:CSR2} then shows that what enters in
the CSR is  just the above expression, however with the T-product
replaced by its equal-time specification and it is precisely this
quantity which is directly connected with the pair distribution
function. As we have seen (and as we will discuss in more detail
later) the latter crucially affects the CSR, namely, the frequency
integral of the response.

The response, however, is not directly expressible in terms of the
pair distribution  correlation function, but rather the momentum
distribution of the system's constituents (in our case a
$\theta$-function) enters into its definition. It thus appears that
the $\theta$-function should be viewed as the fundamental ingredient
of both the system's response and of the CSR.   Indeed in
Sec.~\ref{sec:modif} we shall illustrate how the pair distribution
function (and hence the CSR) is determined by the momentum
distribution.

To proceed further, observe that with a straightforward application
of Wick's theorem to Eq.~\eqref{eq:Pi2} one obtains for the
non-interacting Fermi system
\begin{equation}
\label{eq:Pi3}
\Pi^0(x,0)=-2iG^0(x,0)G^0(x,0) ~.
\end{equation}
To compute this we start from the Fourier transform of its  well-known
expression in energy-momentum space
\begin{eqnarray}
\label{eq:Pi4}
\Pi^0(q)&=&\big(-2 i \big)\int \frac{d^4 k}{(2 \pi)^4} G^0(k) G^0 (k+q)
\nonumber\\
&=&  2 \int \frac{d \vec{k}}{(2 \pi)^3}\theta(|\vec {k} + \vec{q}| -
k_F)  \theta(k_F - k) \bigg [ \frac{1}{q_0 +
\omega_{\vec{k}}-\omega_{\vec{q}+\vec{k}}+ i \eta} -\frac{1}{q_0-
\omega_{\vec{k}}+\omega_{\vec{q}+\vec{k}}- i \eta} \bigg ]\ ,
\nonumber\\
\end{eqnarray}
where the integration on $k^0$ has been performed in the complex
plane. Next  we take the inverse Fourier transform of
Eq.~\eqref{eq:Pi4}
\begin{equation}
\label{eq:Pi5}
\Pi^0(x,0)=\int \frac{d^4 q}{(2 \pi)^4} e^{i q\cdot x}\Pi^0(q)
\end{equation}
and carry out the $q_0$-integration in the complex plane: this again
distinguishes between forward and backward time propagation,
corresponding to the two terms generated by the T-product. Both of
these diverge for $t_x= 0$ in accord with what was previously found
for the one-fermion propagator.  Here we recall that only the
particle piece of the latter was seen to diverge at equal times. We
shall return on this point later on.

Furthermore, the second term of $\Pi^0(x,0)$, associated with $q_0<
0$,  describes,  the system's response in the time-like domain. One
obtains
\begin{eqnarray}
\label{eq:Pi6} \Pi^0(x,0)&=& -2i \int \frac{d \vec{q}}{(2 \pi)^3}
e^{i \vec{q} \cdot \vec{x}} \int \frac{d \vec{k}}{(2 \pi)^3}
\theta(k_F - k) \theta(|\vec {k} +\vec{q}| - k_F)  \nonumber\\ &&
\times  \bigg \{ e^{-i \big ( \frac{q^2}{2 m} + \frac{1}{m}\vec{q}
\cdot \vec{k} \big ) t_x }  \theta (t_x)+ e^{i \big ( \frac{q^2}{2
m} + \frac{1}{m} \vec{q} \cdot \vec{k} \big )t_x } \theta (- t_x)
\bigg \} ~.
\end{eqnarray}
After some algebra (see Appendix \ref{sec:app_pol} for details) the
above  expression can be reduced to a single integral (we focus on
the first piece, since the second one is immediately derived when
the first is known):
\begin{eqnarray}
\label{eq:Pi9}
&& \Pi^0(x,0) \theta(t_x) =
-\frac{9 i n_0^2}{4 \Delta_t k_F r} \bigg \{ \int_{2}^{\infty} ds e^{-i \Delta_t s^2}\sin(s k_F r)
j_{1}(2\Delta_t s)\nonumber\\
&& + \int_{1}^{2} d s e^{-i \Delta_t s^2}\sin(s k_F r) j_{1}(2\Delta_t s(s -1))(s - 1)^2
\nonumber\\
&& +\frac{1}{2\Delta_t} \int_{1}^{2 } ds \sin (s k_F r)\bigg [ e^{-i
\Delta_t s^2}  \cos (2\Delta_t s (s - 1))  -e^{-2 i \Delta_t s^2} \bigg
( \frac{1}{s} \cos \Big ( \Delta_t s (s - 2)
\Big )
\nonumber\\
&& + \frac{1}{2\Delta_t s^2} \sin \Big ( \Delta_t s (s-2) \Big )  \bigg ) \bigg ]+
\frac{1}{2\Delta_t} \int_{0}^{1} ds
\sin(s k_F r )e^{-2 i \Delta_t s} \sin(2\Delta_t s^2) \frac{1}{s}
\bigg (\frac{1}{2\Delta_t s} + i \bigg ) \nonumber\\
&& + \frac{\cos(2 k_F r)-1}{4\Delta_t k_F r } \bigg \}        ~,
\end{eqnarray}
where $r\equiv |\vec x|$  and $s=q/k_F$. As previously anticipated,
in the limit of vanishing $\Delta_t$ the above becomes purely
imaginary and diverges.

To show that this divergence is as severe as the one encountered for
the one-particle propagator (see Eq.~\eqref{eq:int}) we consider the
first term inside the curly brackets. Here we are allowed to replace
the spherical Bessel function with his leading term in the small
argument expansion getting

\begin{equation}
\label{eq:Pi9bis}
\frac{2\Delta_t}{3} \int_{2}^{\infty} ds e^{- i \Delta_t s^2} \sin(s k_F r)
\simeq
\frac{1}{2} \sqrt{\frac{\pi}{\Delta_t i^3}}\frac{k_F r}{3}
e^{i \frac{(k_F r)^2}{4 \Delta_t}} ~.
\end{equation}
Hence our statement follows.

\begin{figure}[htbp]
\centering%
\subfigure[\label{fig:repai08}]%
{\includegraphics[height=4cm]{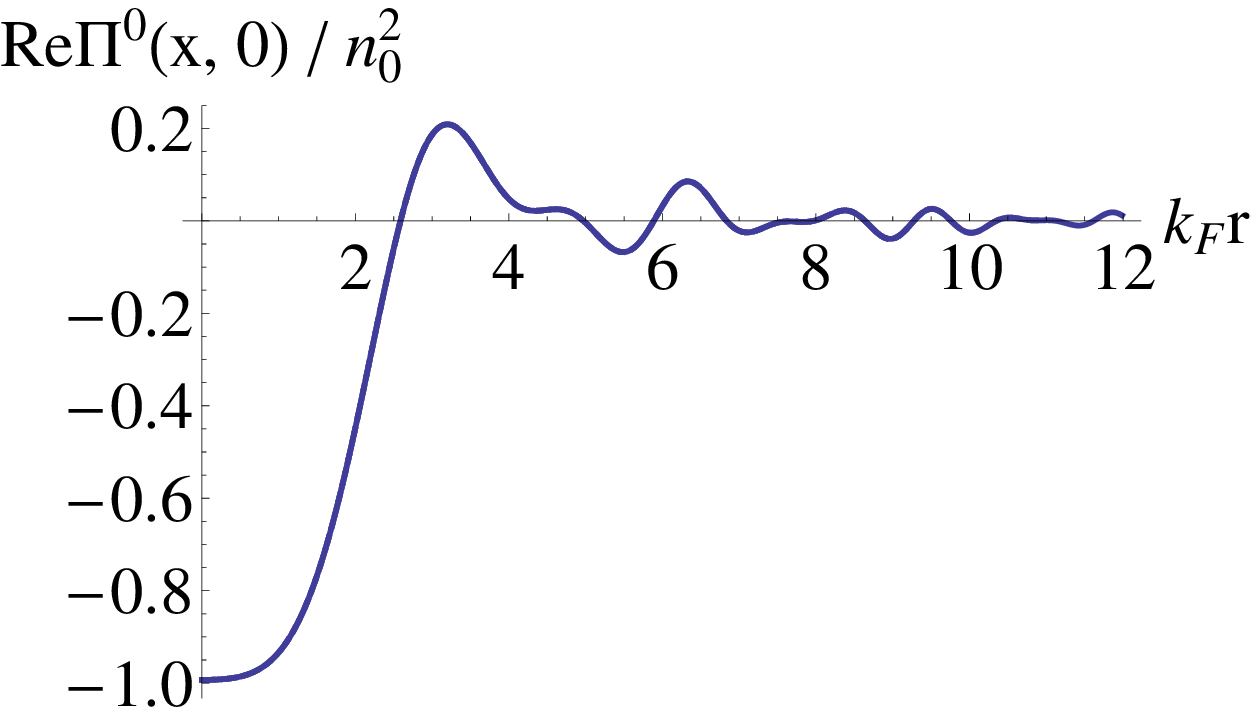}}\qquad\qquad
\subfigure[\label{fig:impai08}]%
{\includegraphics[height=4cm]{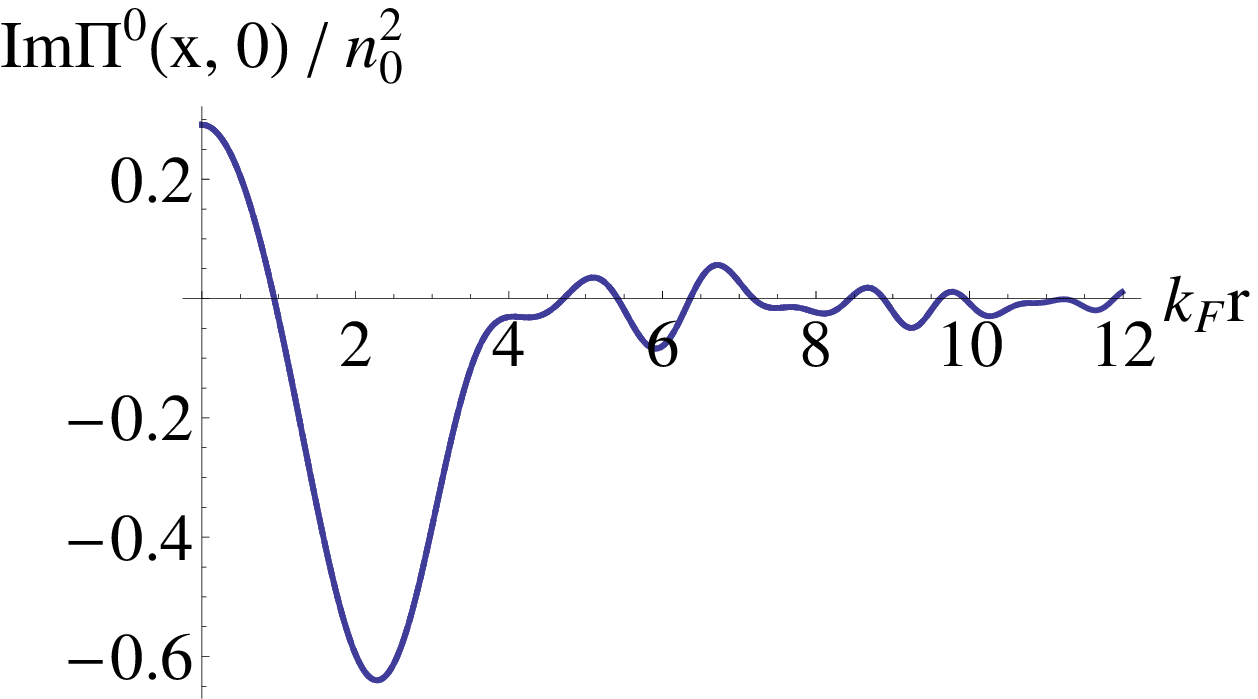}}\qquad\qquad
\subfigure[\label{fig:abspai08}]%
{\includegraphics[height=4cm]{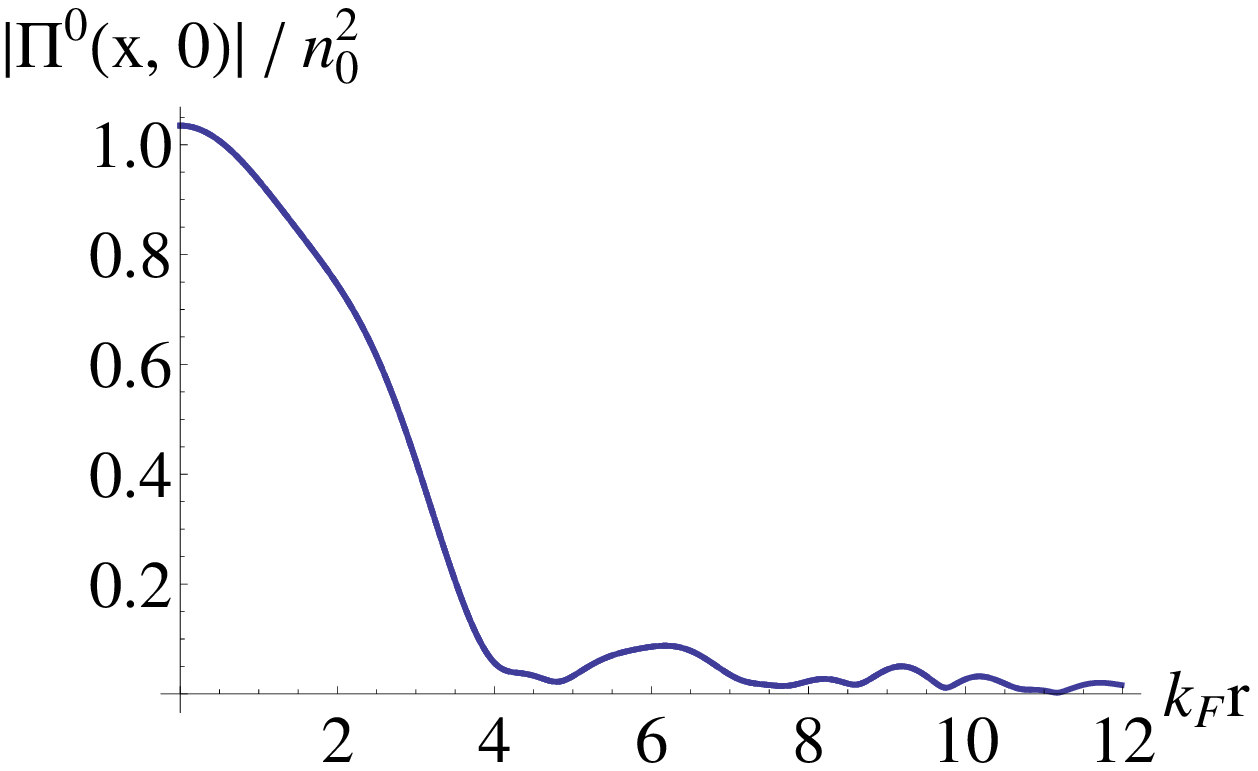}} \caption{
\label{fig:pai08} Real part (a), imaginary part (b) and modulus (c)
of $\Pi^0(x,0)$ divided by $n_0^2$ and plotted versus $k_F r$ for $\Delta_t = 0.8 $
and three values of $k_F$: 1.2, 1.36 and 1.5 fm$^{-1}$.
Note that the same curve is obtained for any value of $k_F$.}
\end{figure}

\begin{figure}[htbp]
\centering%
\subfigure[\label{fig:repai1}]%
{\includegraphics[height=4cm]{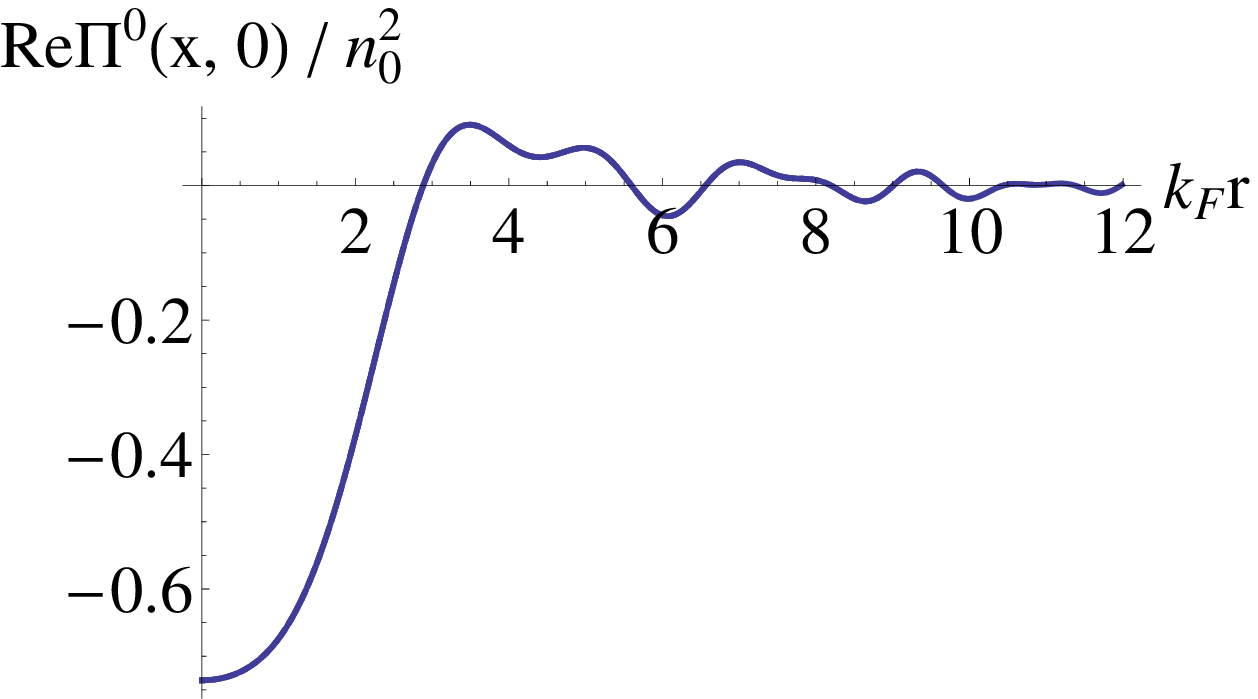}}\qquad\qquad
\subfigure[\label{fig:impai1}]%
{\includegraphics[height=4cm]{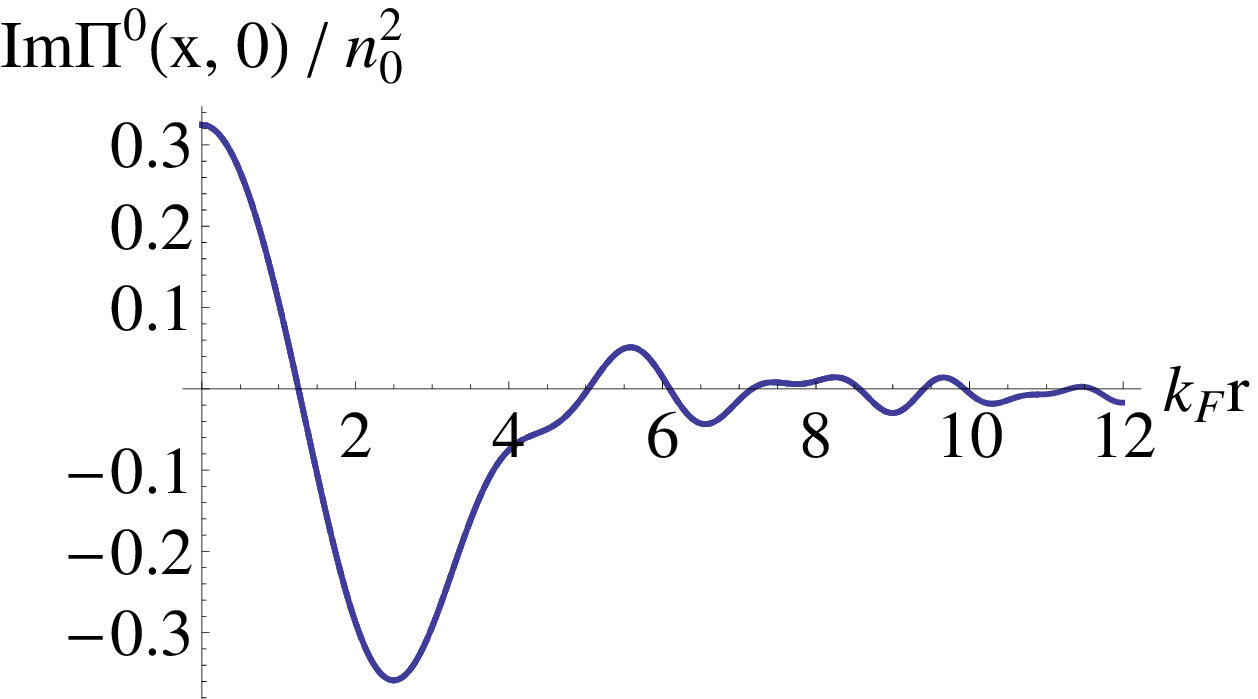}}\qquad\qquad
\subfigure[\label{fig:abspai1}]%
{\includegraphics[height=4cm]{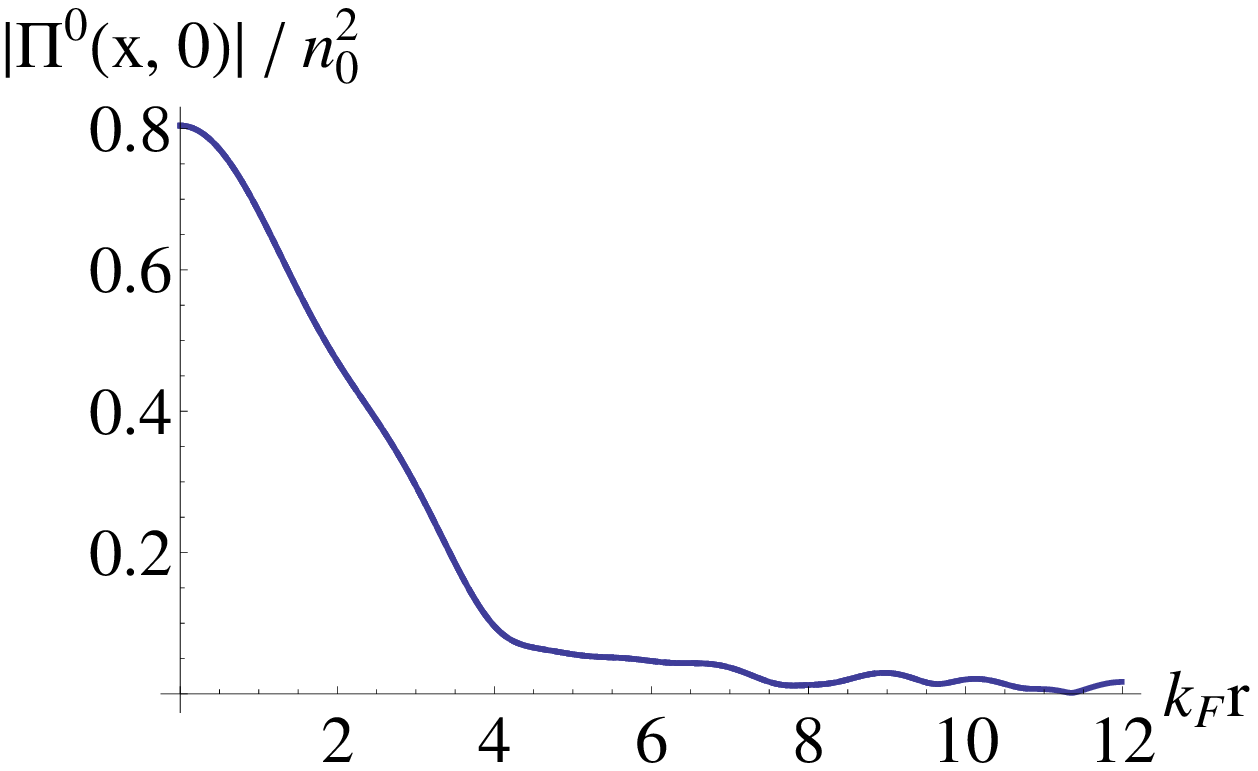}} \caption{
\label{fig:pai1} As for Fig.~\ref{fig:pai08}, but now with $\Delta_t =
1 $~.}
\end{figure}

\begin{figure}[htbp]
\centering%
\subfigure[\label{fig:repai10}]%
{\includegraphics[height=4cm]{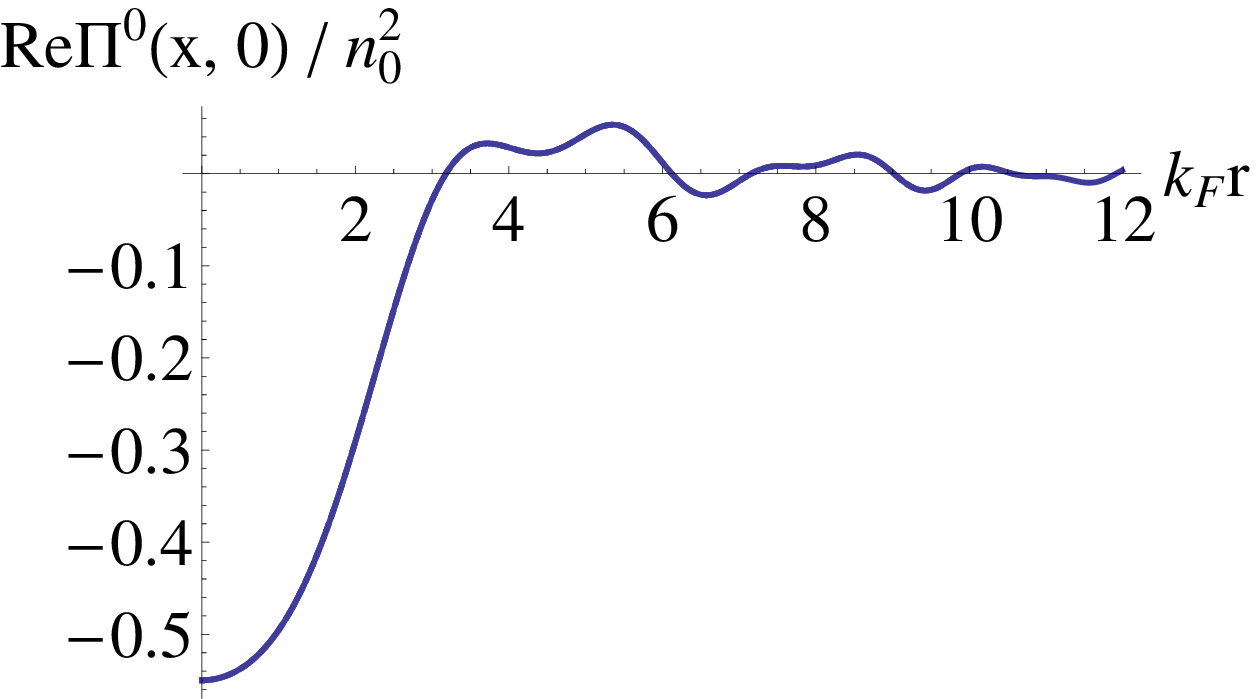}}\qquad\qquad
\subfigure[\label{fig:impai10}]%
{\includegraphics[height=4cm]{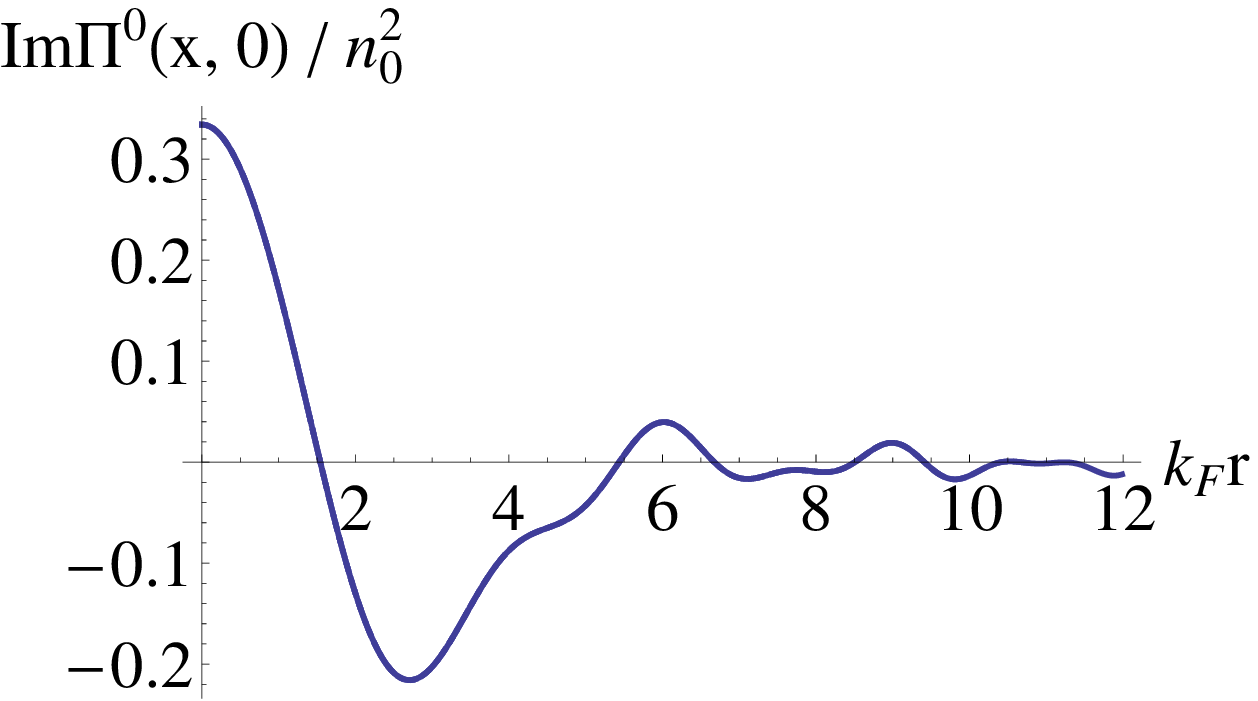}}\qquad\qquad
\subfigure[\label{fig:abspai10}]%
{\includegraphics[height=4cm]{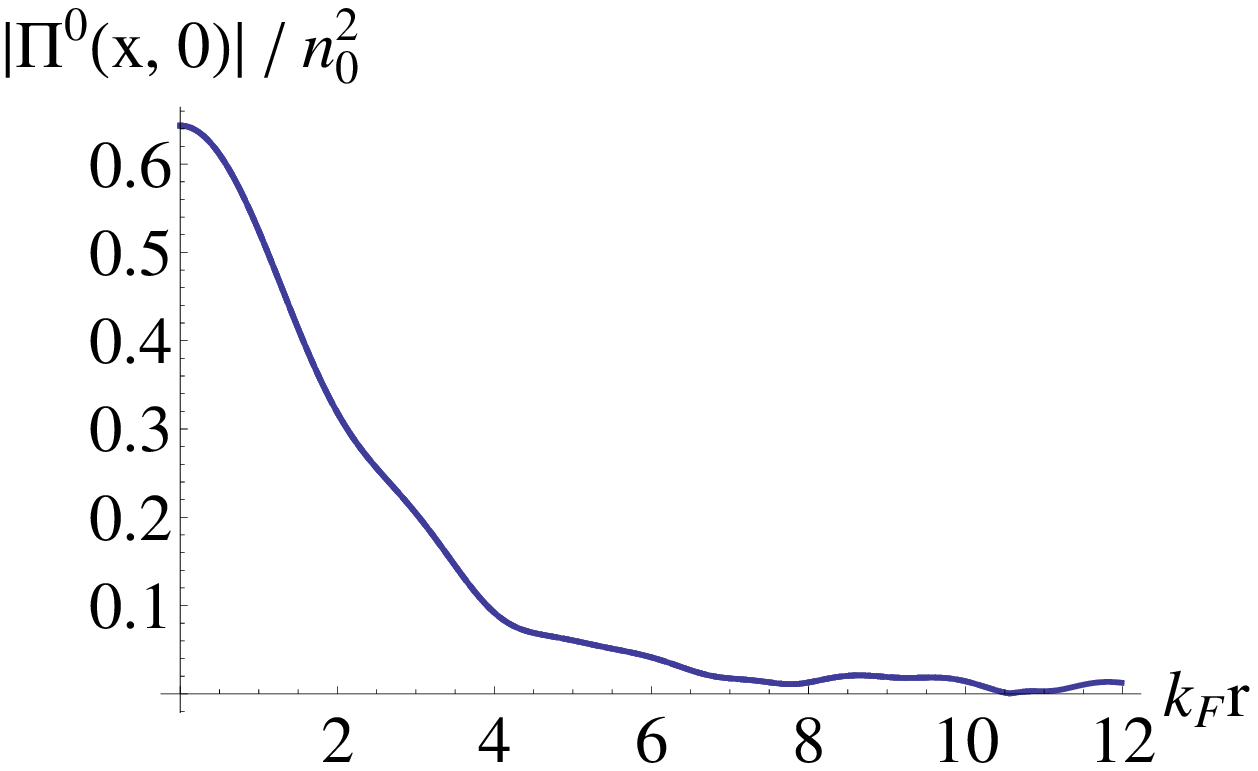}} \caption{
\label{fig:pai12} As for Fig.~\ref{fig:pai1}, but now with $\Delta_t =
1.2 $~.}
\end{figure}

In Figs.~\ref{fig:pai08} -~\ref{fig:pai12}  we display the modulus
of $\Pi^0 (x ,0)$ together with its real and imaginary parts for a
few values of $\Delta_t$. We observe that for small  $\Delta_t$ a
diffraction pattern emerges as in the case of the single-particle
propagator: indeed, from the Heisenberg principle, here the energy
is considerably spread out and, as a consequence, a wave packet can
be set up which vanishes at fixed positions selected by the medium.
As $\Delta_t$ increases this pattern is washed out until for very
large $\Delta_t$ it turns into an almost uniform behaviour. From the
formal point of view this evolution reflects the fact that for small
time differences the particle-hole propagator is essentially
imaginary, just as happened for the one-hole propagator. This
imaginary part oscillates with the distance and hence the
diffraction pattern follows. However, as the time difference
increases a real part develops, also with an oscillatory behaviour,
but with zeros displaced with respect to those of the imaginary
part. Hence the zeros of the diffraction pattern are lifted up until
at large $\Delta_t$ the modulus of $\Pi^0 (x ,0)$ becomes uniform in
space.

A further important feature of $\Pi^0 (x ,0)$ relates to its rapid
decrease occurring in the range $0 < k_F r\lesssim 4$: we have numerically
checked that in this domain all the terms in
Eq.~\eqref{eq:Pi9} contribute by roughly the same amount whereas
for larger $k_F r$ the integrals with the variable running in the
intervals $(0,1)$ and $(1,2)$, where the Pauli correlations are
operative, essentially drop out.
We thus conclude that are these correlations which damp the propagation
of a density disturbance in the system.

We turn now to the evaluation of the remaining integrals with
respect to the variable $q$ to  complete our task of obtaining an
analytic expression for $\Pi^0 (x, 0)$. For this purpose we
introduce the dimensionless quantities
\begin{equation}
\label{eq:Pi10}
z = \frac{k_F r}{2\sqrt{\Delta_t}}  \mbox{   and   }   \rho = \sqrt{\Delta_t}
\end{equation}
and the function
\begin{equation}
\label{eq:Pi11}
g(a,b,c) = \int_a^b dy \frac{e^{i y^2}}{y+c}  ~.
\end{equation}
The latter can in fact be expressed in terms of the Meijer-$G$
functions~\cite{Abr},  although the resulting formulas are  quite
cumbersome and hence we prefer to use directly the definition
expressed by Eq.~\eqref{eq:Pi11}. Even so the remaining
$q$-integrations are, unfortunately, given by expressions which are
far from simple;  we report these in Appendix \ref{sec:app_pol} for
completeness.

To pave the way  to Appendix \ref{sec:app_pol} here we simply
rewrite Eq.~\eqref{eq:Pi9} in terms of the variables in
Eq.~\eqref{eq:Pi10}. It becomes
\begin{eqnarray}
\Pi^0_a(x,x') &=& -i n_0^2 \frac{9}{8 z} \frac{1}{\rho^3}
\int_2^\infty ds e^{-i\rho^2 s^2} \sin(2\rho z s) j_1(2\rho^2 s)
\nonumber
\\
\label{eq:baia}
&\simeq& -i n_0^2 \frac{3}{2} \int_2^\infty ds e^{-i\rho^2 s^2} s^2 \,\,\, \mbox{\rm for } \rho\to 0
\end{eqnarray}
\begin{eqnarray}
\Pi^0_b(x,x') &=& -i n_0^2 \frac{9}{8 z} \frac{1}{\rho^3}
\int_1^2 ds e^{-i\rho^2 s^2} \sin(2\rho z s) j_1(2\rho^2 s(s-1)) (s-1)^2
\nonumber
\\
\label{eq:baib}
&\simeq& -i n_0^2 \frac{49}{40} \,\,\, \mbox{\rm for } \rho\to 0
\end{eqnarray}
\begin{eqnarray}
\Pi^0_c(x,x') &=& -i n_0^2 \frac{9}{16 z} \frac{1}{\rho^5}
\int_1^2 ds \sin(2\rho s z) \left\{
e^{-i\rho^2 s^2} \cos(2\rho s(1-s)) -
e^{-2 i\rho^2 s^2}
\right.
\nonumber\\
&\times&\left.\left[ \cos(\rho^2 s(s-2)) + \frac{1}{2\rho^2 s} \sin(\rho^2 s(s-2))\right]
\right\}
\nonumber
\\
&\simeq& -i n_0^2 \frac{21}{16} \frac{1}{\rho^4} \,\,\, \mbox{\rm for } \rho\to 0
\label{eq:baic}
\end{eqnarray}
\begin{eqnarray}
\Pi^0_d(x,x') &=& -i n_0^2 \frac{9}{16 z} \frac{1}{\rho^5}
\left\{
\frac{\cos(4\rho z)-1}{8 z \rho^3} k_F^3 +
\int_0^1 ds \sin(2\rho s z)
e^{-2 i\rho^2 s} \sin(2\rho^2 s^2) \frac{1}{s} \left(\frac{1}{2 s\rho^2}+i\right)
\right\}
\nonumber\\
&\simeq& i n_0^2 \frac{9}{8} \frac{z}{\rho^4} \,\,\, \mbox{\rm for } \rho\to 0
\label{eq:baid}
\end{eqnarray}

Concerning this all important issue, referred to as \textit{second
kind scaling}, much light on it is shed by the analysis of
$\Pi^0(x,0)$ in space-time coordinates. Here one realizes that just
as $G^0(x,x')$ turned out to be proportional to $n_0$ (see
Eq.~\eqref{eq:G04}), $\Pi^0(x,0)$ is proportional to $n_0^{2}$ as
expected. Then, when these density factors are divided out in both
propagators one finds that $G^0(x,x')$  and $\Pi^0(x,0)$ scale at
any $\Delta_t$ providing that the space-time coordinates are in turn
rescaled in terms of the Fermi momentum and frequency. This is
clearly illustrated in Figs.~\ref{fig:pai08}, \ref{fig:pai1}
 and \ref{fig:pai12}
where the spatial behaviour of  the $\Pi^0(x,0)$ associated with
three different values of $k_F$ is displayed as a function of $k_F
r$. The curves are seen to coincide at $\Delta_t =0.8$, $1$ and
$1.2$, in accord with Eq.~\eqref{eq:Pi9}, which transparently
exhibits the $k_F$-scaling property.

The above results concerning  $\Pi^0(x,0)$ should, however, be
viewed with some care. This is because the search for scaling in
$k_F$ at vanishing $\Delta_t$ is impossible owing to the divergence
discussed above which signals that the theory is ill-defined.
Actually all of the terms in Eqs.~(\ref{eq:Pi9}) diverge at
$\Delta_t = 0$, with the exception of the second one.

\section{The impact of a more realistic momentum distribution}
\label{sec:modif}

Here we explore the impact of interactions among the constituent
fermions on the pair correlation function and the CSR. This we do in
a schematic frame that should, however, capture some of the relevant
physics. We assume for the momentum distribution $n(k)$ the
expression
\begin{eqnarray}
\label{eq:B1} n(k)= \theta (k_F - k ) (1- \alpha \frac{k^2}{k_F^2}
)+ \theta (k - k_F)    \beta_1 e^{-\beta_2 (\frac{k}{k_F} -1)} ~.
\end{eqnarray}
The four parameters (indeed also $k_F$ should be viewed as such)
entering in Eq.~\eqref{eq:B1}  must satisfy the normalization
condition
\begin{eqnarray}
\label{eq:B2} \frac{k_F^3}{\pi^2} \bigg [ \frac{1}{3}
-\frac{\alpha}{5} + \frac{\beta_1}{\beta_2^3}  (\beta_2^2 +2
\beta_2+2) \bigg ] =  n_0\ ,
\end{eqnarray}
where $n_0$ is the \textit{experimental} constant density of the
system. Notice that for $1- \alpha =  \beta_1$ the Fermi system
becomes superconductive  (the Fermi surface disappears) whereas for
$1-\alpha > \beta_1$ the discontinuity at the Fermi surface remains,
as it should for a normal Fermi system according  to the Luttinger~\cite{Lutt}
theorem. This implies that both $\alpha$ and $\beta_1$ should be
positive.

\begin{figure}[htbp]
\centering
\includegraphics[height=5cm]{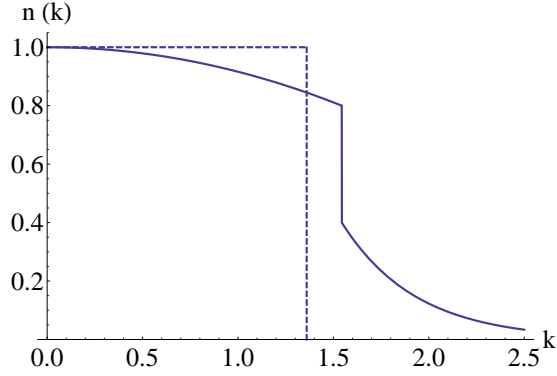}
\caption{\label{fig:nk}  The momentum distribution of an interacting
Fermi system as given by formula Eq.~\eqref{eq:B1} of the text. The
following values for the parameters have been chosen $k_F = 1.54
$ fm$^{-1}$, $\alpha = 0.2$ , $\beta_1 = 0.4$ and $\beta_2 = 4$~. The
non interacting case (dotted line) corresponds to $k_F = 1.36
$ fm$^{-1}$~. }
\end{figure}

We now choose as an example nuclear matter (in this case the
left-hand side of Eq.~\eqref{eq:B2}  should be multiplied by $2$ to
account for the isospin degeneracy) where one has $n_0 = 0.17$
fm$^{-3}$. We then display in Fig.~\ref{fig:nk} the $n (k)$ for
nuclear matter for a specific choice for the four parameters, chosen
to fulfill both Eq.~\eqref{eq:B2} and the above-mentioned
constraint. The tail at large momenta is evident and one finds that
the new Fermi momentum turns out to be $k_F = 1.54 $ fm$^{-1}$,
namely \textit{larger} that the one associated with the
non-interacting case.

\begin{figure}[htbp]
\centering
\includegraphics[height=5cm]{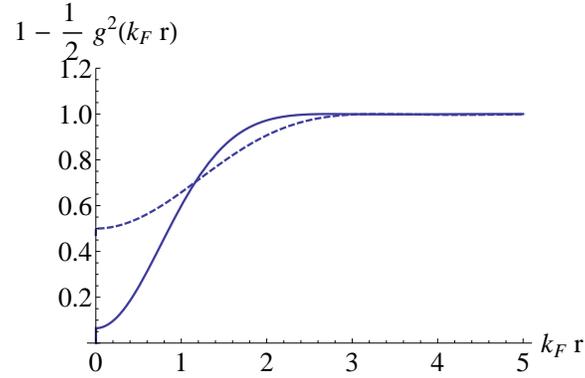}
\caption{\label{fig:nk1} The pair correlation function $1 -
\frac{1}{2} g^2 (k_F r)$ for a  free (dotted line) and for an
interacting (continuos line) Fermi gas as given by our model
(formula ~\eqref{eq:B3} of the text). The values of parameters are
the same as in Fig.~\ref{fig:nk}~. Note that the two curves refer to
different $k_F$~. }
\end{figure}

The pair correlation function for the momentum distribution in
Eq.~\eqref{eq:B1}  is then easily computed and reads
\begin{eqnarray}
\label{eq:B3} && <\Psi_0|\hat{\Psi}_{\alpha}^{\dagger}(\vec
x)\hat{\Psi}_{\beta}^{\dagger}(\vec 0)  \hat{\Psi}_{\beta}(\vec 0)
\hat{\Psi}_{\alpha} (\vec x)|\Psi_0>=n_0^2 \Big \{ 1 -
\frac{1}{2}g^2 (r) \Big \}
\nonumber\\
&& = n_0^2 \Bigg \{ 1 - \frac{1}{2} \bigg \{ \frac{3}{k_F r} \bigg [
j_1(k_F r) -  \frac{\alpha}{(k_F r)^4} \Big ( 3 ((k_F r)^2 -2)
\sin(k_F r)
\nonumber\\
&& - k_F r ((k_F r)^2 -6) \cos(k_F r) \Big ) + \beta_1 \Big( \frac{k_F r} {(k_F r)^2 +\beta_2^2} \Big )^2  \\
&& \times \Big ( \sin (k_F r) \Big ( \beta_2 + \frac{\beta_2^2+
\beta_2^3}{(k_F r)^2} -1 \Big ) +\cos (k_F r) \Big ( k_F r +\frac{2
\beta_2 +  \beta_2^2}{k_F r} \Big ) \Big )\bigg ]
 \bigg \}^2 \Bigg \} \nonumber  ~.
\end{eqnarray}
Using the same values of the parameters as for the momentum
distribution shown in Fig.~\ref{fig:nk}, this pair correlation
function is displayed in Fig.~\ref{fig:nk1} where it is compared
with that of the pure Fermi gas from Eq.~\eqref{eq:corrf}. What
clearly appears in the figure is the marked difference between the
two correlations functions at short distances, while they
practically coincide at large distances: this behaviour nicely
illustrates the role of the short-range correlations.

Finally, we compute the CSR using Eq.~\eqref{eq:B3}. Although also
in this case  the calculation can be analytically performed using
complex coordinates, as was previously done in the non-interacting
situation, the resulting expression  turns out to be very
cumbersome; hence we resort to the numerical evaluation of the
formula
\begin{eqnarray}
\label{eq:B4} S(\vec q) = Z - n_0^2 \frac{1}{2} \int d \vec{r} e^{-i
\vec{q} \cdot \vec{r}} g^2(r)\ ,
\end{eqnarray}
the function $g(r)$ being defined in Eq.~\eqref{eq:B3}.

\begin{figure}[htbp]
\centering
\includegraphics[height=5cm]{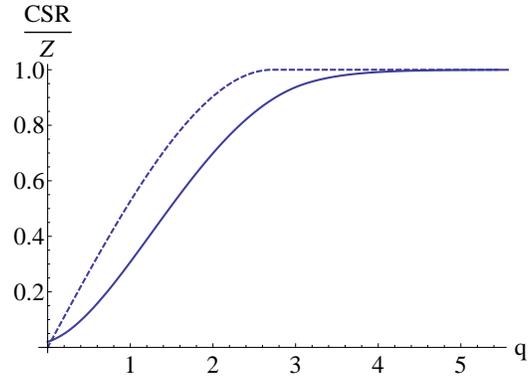}
\caption{\label{fig:csr}  The Coulomb sum rule (see
Eq.~\eqref{eq:B4} of the text) for a free Fermi gas (dotted line)
and for a correlated one according to our model (continuous line).
The parameters are the same as in Fig.~\ref{fig:nk}  and
Fig.~\ref{fig:nk1}~.}
\end{figure}

The outcome is shown in Fig.~\ref{fig:csr} where it is clearly seen
that results from Eq.~\eqref{eq:B4}  coincide with those from
Eq.~\eqref{eq:CSR4} at large $q$ (say $q > 4$ fm$^{-1}$), as they
should, since in this domain of momenta the associated wavelengths
are so small that the system appears to the probe as a collection of
uncorrelated fermions. On the other hand for, say, $ 2.5 < q < 4$
fm$^{-1}$ the two sum rules differ substantially due to the action
of the correlations among the fermions. Finally, for smaller $q$
this difference tends to disappear  as both sum rules should go to
zero when $q$ vanishes. Noteworthy is that the sum rule arising from
Eq.~\eqref{eq:B4} reaches the asymptotic value 1 from below.

\section{Conclusions}
\label{sec:concl}

In this work we have deduced expressions for the one- and
two-particle propagators as functions of space-time coordinates,
focusing on the hole propagator and the density-density correlator.
To our knowledge these expressions were not previously available in
the literature. We find that both propagators have infinities at
equal time, a problem that is being addressed in other work. Next we
have explored the asymptotic space-time behaviour of both Green
functions and have found a transition from a diffractive regime to a
uniform one as the time difference between the fields (for the
one-particle propagator) or between the densities (for the
two-particle propagator) grows. From the formal point of view this
relates to the fact that both propagators for zero time difference
are purely imaginary, but then the real parts start to develop as
the time difference grows. Concerning the dependence upon $k_F$, our
analysis shows that both $G^0$ and $\Pi^0$ scale once appropriate
measures for space and time are chosen. This outcome goes in
parallel with the situation in frequency-momentum space. However the
divergence affecting $\Pi^0$ at $\Delta_t=0$ prevents the analysis
of second-kind scaling for this propagator at vanishing $\Delta_t$.

We have found that the key ingredient contained in the propagators
is the momentum distribution $n(k)$. For example the Coulomb sum
rule can be directly expressed in terms of the pair correlation
function, which can, of course, be obtained once $n (k)$ is known.
We have explored the consequences arising from a modification around
the Fermi surface of $n (k)$ away from the pure Fermi gas result,
finding that this induces striking effects in the pair correlation
function at short distances. This in turn leads to major differences
in the Coulomb sum rule for momenta between about $2.5$ and $4$
fm$^{-1}$ for the model used in the present study, suggesting that
it would be interesting to explore the responses of our infinite
Fermi system to external probes, employing in the calculation of the
polarization propagator our modified momentum distribution function
given in Eq.~\eqref{eq:B1}. This, of course, is meant to account for
the correlations (possibly of short range) among the fermions.

A  further issue, in some sense complementary to the above one,
deserves consideration. It relates to the passage from an infinite
to a finite system, the latter obviously of concern for nuclear
physics. As a first approximation this transition can be
accomplished by accounting for the modification of the density of
the single-particle states induced by the presence of the system's
surface according to the prescription given by Feshbach \cite{Fesh}.
In a preliminary investigation we have done so and found, as
expected, that the impact of the surface in coordinate space on the
system's response is only felt at low momenta. More specifically,
the confinement on the one hand entails oscillations of the system's
density in coordinate space, on the other enlarges the momentum
distribution $n(k)$ to momenta greater than those of the
corresponding infinite system with equal density. At the same time
it digs a hole at low $k$ in the momentum distribution. Thus
confinement (in leading order) and short-range correlations appear
to work in the same direction at large momenta.
 The disentangling of the interplay
between the two effects is a problem we intend to address in
forthcoming work. We will also explore in depth how the scaling of
first kind is reflected in the space-time coordinates.

Finally, since much of the physics we are addressing here occurs at
large momenta, it is imperative to extend the present treatment to
the relativistic context, with the goal of providing a comparison
between the results one would obtain using the present approach and
those already obtained in other approaches to electroweak
superscaling. For instance, it will be of interest to answer the
question: How will scaling in $k_F$ of the imaginary part of the
polarization propagator occurring for all values of space-time
coordinates in the non-interacting, homogeneous case be affected
(and eventually disrupted) by confinement and short-range
correlations in both the non-relativistic and relativistic context?

\begin{acknowledgments}
We like to thank Prof. G. Chanfray and Dr. H. Hansen for useful
discussions.
This work was partially supported (TWD) by
U.S. Department of Energy under cooperative
agreement DE-FC02-94ER40818.
\end{acknowledgments}

\appendix

\section{}
\label{sec:app_prop}

In this appendix we analyze the behaviour of the propagator
$G^0(x,x')$  for very small $\Delta_t$. For this purpose we use the
well-known  asymptotic expansion of the error function
\begin{eqnarray}
\label{eq:A1} erf(z) = 1 - \frac{1}{\sqrt{\pi} z} e^{-z^2} \Bigg [1+
\sum_{m=1}^{\infty}  (-1)^m \frac{(2 m -1)!!}{(2 z^2)^m} \Bigg ]~.
\end{eqnarray}
Then from the above, after  some algebra, in leading order of
$\Delta_t$ one  obtains for the third term on the right-hand side of
the propagator in Eq.~\eqref{eq:G04}
\begin{eqnarray}
\label{eq:A2} && \frac{1}{2} \frac{k_F|{\vec{ x}}-{\vec
{x}'}|}{2\Delta_t }   \bigg ( \frac{e^{i k_F |{\vec{ x}}-{\vec
{x}'}| }}{k_F |{\vec{ x}}-{\vec {x}'}| + 2 \Delta_t} -\frac{e^{- i
k_F |{\vec{x}}-{\vec {x}'}| }}{k_F |{\vec{ x}}-{\vec {x}'}| - 2
\Delta_t} \bigg)
\nonumber\\
&& =  - \frac{1}{2 i \Delta_t} \sin(k_F |{\vec{ x}}-{\vec {x}'}|)
-\frac{\cos(k_F |{\vec{ x}}-{\vec {x}'}|)}{k_F |{\vec{
x}}-{\vec{x}'}|}
\end{eqnarray}
which  exactly cancels the second term of the right-hand side;
hence Eq.~\eqref{eq:G0equalt} follows.

\section{}
\label{sec:app_integral}

Here we compute the integral in Eq.~\eqref{eq:CSR3} working in the
complex $z$-plane. Setting ${\overline q}=q/k_F$, we have
\begin{eqnarray}
\label{eq:I1}
I&=& \int_{0}^{\infty} d z  j_{1}^2(z) j_{0}(\overline q z) = \frac{1}{2} \int_{-\infty}^{\infty} d z \bigg ( \frac{\sin
z}{z^2} - \frac{\cos z }{z} \bigg )^2 \frac{\sin (\overline q  z)}{\overline q z}
\nonumber\\
&=& \frac{1}{2 \overline q} \int_{-\infty}^{\infty}\frac{d z}{z^5} (\sin z - z \cos z)^2 \sin(\overline q z)  ~.
\end{eqnarray}
The integrand, which behaves as $\sim  z^2$ for $z \rightarrow 0$,
is a  regular analytic function; hence the integration path along
the real axis can be deformed by inserting a very small semicircle
going around the origin from below. This we indicate with the symbol
$\landdownint$. Closing the integration path along a large
semicircle in the $Im \; z > 0$ region, we thus get
\begin{eqnarray}
\label{eq:I2} I = \frac{i}{8 \overline q } \landdownint_{-
\infty}^{\infty} \frac{d z}{z^5} &\Big \{& (1-i z)^2 e^{i (2 +
\overline q)z}- (1-i z)^2 e^{i (2 - \overline q)z} + (1+ i z)^2 e^{i
(\overline q - 2 )z}
\nonumber\\
&&- (1 + i z )^2 e^{-i(\overline q + 2)z} - 2 (1+ z^2) e^{i \overline q z} + 2 (1+ z^2) e^{-i \overline q z} \Big \}\ ,
 \end{eqnarray}
where the  fourth and the sixth term on the right-hand side do not
contribute  because for these the integration path has to be closed
in the $Im \; z < 0 $ domain where no singularities exist. It is
then convenient to split Eq.~\eqref{eq:I2} into two pieces according
to
\begin{eqnarray}
\label{eq:I3}
I = I_1 + I_2
\end{eqnarray}
with
\begin{eqnarray}
I_1 = \frac{i}{8 \overline q } \landdownint_{- \infty}^{\infty}
\frac{d z}{z^5} \Big [ (1- i z)^2 e^{i (\overline q +2)z} - 2
(1+z^2) e^{i \overline q z} \Big ]
\end{eqnarray}
and
\begin{eqnarray}
I_2 = \frac{i}{8 \overline q }
\begin{cases}
\landdownint_{- \infty}^{\infty} \frac{d z}{z^5} (1+ i z)^2 e^{i (\overline q -2)z} & \text{if $\overline q > 2 $ }\\
-\landdownint_{- \infty}^{\infty} \frac{d z}{z^5} (1 - i z)^2 e^{i (2 - \overline q)z} & \text{if $\overline q < 2 $}  ~.
\end{cases}
\end{eqnarray}
The straightforward, while somewhat tedious, computation of the
residues  then yields
\begin{eqnarray}
\label{eq:I5}
I_1 = - \frac{\pi}{8 \overline q} \Big \{ \frac{1}{4!}(\overline q +2)^4 - \frac{2}{3!} (\overline q +2)^3 + \frac{1}{2}
(\overline q+2)^2 - \frac{2}{4!} {\overline q}^4 + {\overline q}^2\Big \} ~,
\end{eqnarray}
\begin{eqnarray}
\label{eq:I6} I_2 =- \frac{\pi}{8 \overline q } \Big \{
\frac{1}{4!}(\overline q -2)^4 + \frac{2}{3!} (\overline q - 2)^3  +
\frac{1}{2}(\overline q -2)^2 \Big \} & \text{if $\overline q > 2 $
}
\end{eqnarray}
and
\begin{eqnarray}
\label{eq:I7} I_2 =+ \frac{\pi}{8 \overline q } \Big \{
\frac{1}{4!}(\overline q -2)^4 + \frac{2}{3!} (\overline q - 2)^3  +
\frac{1}{2}(\overline q -2)^2 \Big \} & \text{if $\overline q < 2 $
}  ~.
\end{eqnarray}
As a result of the cancellation  occurring between
Eqs.~\eqref{eq:I5}  and \eqref{eq:I6} one thus finds
\begin{eqnarray}
\label{eq:I8}
I = I_1 + I_2 =0 & \text{if $\overline q > 2 $ }
\end{eqnarray}
and
\begin{eqnarray}
\label{eq:I9} I = I_1 + I_2  = \frac{\pi}{8} \Big ( \frac{4}{3} -
\overline q  + \frac{1}{12} {\overline q}^3 \Big )& \text{if
$\overline q < 2 $ } ~.
\end{eqnarray}

\section{}
\label{sec:app_pol}

In this appendix we perform the integrals in Eq.~\eqref{eq:Pi6}. We
focus on  the first piece, since the second one is immediately
derived once the first is known. For these pieces three angular
integrations are trivial whereas the fourth should be approached
with care owing to the presence of the $\theta$-function, leading
naturally to the splitting of the integral over the modulus of the
vector $\vec{q}$ into three pieces:
\begin{eqnarray}
\label{eq:Pi7}
&&\Pi^0(x,0) \theta (t_x) =-\frac{4i}{r}\int_{0}^{k_F} \frac{d k}{(2 \pi)^2} k^2
\nonumber\\
&& \times \bigg \{ \int_{2 k_F}^{\infty} \frac{d q }{(2 \pi)^2} q
e^{-i \frac{q^2 }{2 m}t_x} \sin(q r ) \int_{-1}^{1} d\cos\theta
e^{-i \frac{q k \cos\theta}{m} t_x}
\nonumber\\
&& + \int_{k_F}^{2 k_F} \frac{d q }{(2 \pi)^2} q e^{-i \frac{q^2}{2
m}t_x } \sin(q r ) \int_{max
\big[\frac{k_{F}^2-q^2-k^2}{2 q k},-1 \big]}^{1} d\cos\theta e^{-i \frac{q k \cos\theta}{m} t_x} \\
&& + \int_{0}^{k_F} \frac{d q }{(2 \pi)^2} q e^{-i \frac{ q^2 }{2m}
t_x } \sin(q r ) \int_{\frac{k_{F}^2-q^2-k^2}{2 q k}}^{1}
d\cos\theta e^{-i \frac{q k \cos\theta}{m} t_x } \theta \bigg(
1-\frac{k_{F}^2-q^2-k^2}{2 q k}\bigg) \bigg \} \nonumber  ~.
\end{eqnarray}
The remaining angular integration is trivial and the integral over
the  modulus of the vector $\vec{k}$, while somewhat cumbersome, can
be performed. Introducing for convenience a time variable with the
dimensions of a length squared, $\tau = t_x/m$, one arrives at the expression
\begin{eqnarray}
\label{eq:Pi9ter}
&&\Pi^0(x,0)\theta (t_x)=-\frac{i}{2\pi^4 \tau r} \bigg \{ k_F^2 \int_{2 k_F}^{\infty} dq
e^{-i \frac{\tau}{2} q^2}\sin(q r) j_{1}(\tau q k_F)\nonumber\\
&& + \int_{k_F}^{2 k_F} d q e^{-i \frac{\tau}{2} q^2}\sin(q r) j_{1}(\tau q(q -k_F))(q - k_F)^2
\nonumber\\
&& +\frac{1}{\tau} \int_{k_F}^{2 k_F} dq \sin (q r)\bigg [ e^{-i \frac{\tau}{2} q^2} \cos (\tau q (k_F - q))  -e^{-i
\tau q^2} \bigg ( \frac{k_F}{q} \cos \Big ( \tau \Big( \frac{q^2}{2} - q k_F \Big ) \Big )
\nonumber\\
&& + \frac{1}{\tau q^2} \sin \Big ( \tau \Big ( \frac{q^2}{2} - q
k_F \Big )  \Big ) \bigg ) \bigg ]+ \frac{1}{\tau} \int_{0}^{k_F} d
q
\sin(q r )e^{-i \tau q k_F} \sin(\tau q^2) \bigg (\frac{1}{\tau q^2} + i \frac{k_F}{q} \bigg )  \nonumber\\
&& + \frac{\cos(2 k_F r)-1}{2\tau r } \bigg \}        ~.
\end{eqnarray}
Then, for the first term in Eq.~\eqref{eq:Pi9ter}, the
$q$-integration between $2 k_F$ and $\infty$ yields:
\begin{eqnarray}
\label{eq:Pi12} && \Pi^0_{(1)}(x,0) \theta (t_x)  = - \frac{2}{(2
\pi)^4} \frac{\theta (t_x)}{\tau^3} \frac{1}{z} \Big \{  \sqrt{2
\pi} (1-i) e^{-2 i \rho z} \cos(\rho^2 + z^2)
\\
&& + (-2 i \rho \sin (\rho +z)^2 + e^{-i (\rho +z)^2}z) g^*
(-\infty,+ \infty,\rho +z) + e^{i (\rho -z)^2} (2 \rho -z)
g^*(-\infty, + \infty, \rho - z ) \Big \} \nonumber   ~;
\end{eqnarray}
for the second term in Eq.~\eqref{eq:Pi9ter} the $q$-integration
between $k_F$ and $2 k_F$ yields
\begin{eqnarray}
\label{eq:Pi13}
&& \Pi^0_{(2a)}(x,0) \theta (t_x) = - \frac{i}{(2 \pi)^4}
\frac{\theta (t_x)}{\tau^3}  \frac{1}{z}  \nonumber\\
&& \times \bigg \{ -  \frac{\sin(4 \rho z)}{\rho} (1 - e^{-i 8
\rho^2}) + 4  \frac{e^{-i \rho^2}}{\rho} \sin(2 \rho z) +
\sqrt{\pi}\frac{1 + i}{2} \Big [ e^{-i (\rho -z)^2} erf
\Big(\frac{1-i}{\sqrt{2}} (\rho + z)\Big)
\nonumber\\
&& - e^{i(\rho +z)^2} erf \Big(\frac{1-i}{\sqrt{2}} (\rho -
z)\Big)-\big ( e^{-i (\rho - z)^2 }+e^{-i (\rho + z)^2 } \big ) erf
\Big(\frac{1-i}{\sqrt{2}} z \Big )
\nonumber\\
&& - i \sqrt{3}  \Big( erf
\Big(\frac{1+i}{\sqrt{2}}\frac{1}{\sqrt{3}} (5 \rho - z)\Big) -  erf
\Big(\frac{1+i}{\sqrt{2}} \frac{1}{\sqrt{3}} (2 \rho - z)\Big) \Big)
e^{\frac{i}{3}(\rho +z)^2}  + (z \rightarrow -z)  \Big ]
\nonumber\\
&& - 2 \Big [ e^{ -i (\rho - z)^2} (\rho - z) g(z, \rho + z, \rho -
z) - e^{i (\rho + z)^2}  (\rho + z) g(-z, \rho - z, \rho +z)
\nonumber\\
&& + \Big ( \frac{1}{\sqrt{3}} e^{\frac{i}{3} (\rho +z)^2} (\rho +z)
g\Big(\frac{3 \rho - z} {\sqrt{3}},\frac{5 \rho -
z}{\sqrt{3}},\frac{ \rho + z}{\sqrt{3}} \Big ) - (z \rightarrow -z)
\Big ) \Big ]
\nonumber\\
&&
 -(1+i) \sqrt{\frac{\pi}{2}} \Big [ \Big ( e^{- i (\rho - z)^2} \Big ( erf \Big(\frac{1-i} {\sqrt{2}}  (\rho + z)\Big) -
erf \Big(\frac{1-i}{\sqrt{2}} z\Big) \Big ) -(z \rightarrow -z)
\nonumber\\
&& + \Big ( \frac{1}{i}  e^{\frac{i}{3} (\rho +z)^2} \Big ( erf
\Big(\frac{1+i}{\sqrt{2}}\frac{1}{\sqrt{3}} (5 \rho - z)\Big) - erf
\Big(\frac{1+i}{\sqrt{2}} \frac{1}{\sqrt{3}} (2 \rho - z)\Big)\Big )
- (z \rightarrow -z) \Big )
\nonumber\\
&& - \frac{2}{1+i} \sqrt{\frac{2}{\pi}} \rho \Big ( \Big
(e^{\frac{i}{3}(\rho +z)^2} g \Big(\frac{2 \rho -
z}{\sqrt{3}},\frac{5 \rho - z}{\sqrt{3}},\frac{ \rho + z}{\sqrt{3}}
\Big )  - (z \rightarrow -z) \Big )
\nonumber\\
&& + \Big ( e^{-i (\rho -z)^2} g(\rho, 2 \rho, \rho - z) - (z
\rightarrow -z) \Big) \Big )  \Big] \Big \}
\end{eqnarray}
for the term embodying the $j_1$ and
\begin{eqnarray}
&& \Pi^0_{(2b)}(x,0) \theta (t_x) = -  \frac{i}{(2 \pi)^4}
\frac{\theta (t_x)}{\tau^3}  \frac{1}{z}  \nonumber\\
&& \times \left\{ e^{\frac{i}{3}(\rho+z)^2}
\left(\frac{1}{\sqrt{3}}-2i\sqrt{3}\right) \sqrt{\frac{\pi}{2}}
\frac{1-i}{2} \left[ erf \left( \frac{1+i}{\sqrt{2}} \left( 2
\sqrt{3} \rho - \frac{\rho + z}{\sqrt{3}} \right) \right)   \right.
\right.
\nonumber\\
&& \left. - erf \left( \frac{1+i}{\sqrt{2}} \left(  \sqrt{3} \rho -
\frac{\rho + z}{\sqrt{3}} \right) \right) \right] - (z \rightarrow
-z)
\nonumber\\
&& - e^{\frac{i}{3}(\rho + z)^2} \left[ \rho \left( 1 - 2 i \right)
- 2 i z \right] g^{*} \left( \sqrt{3} \rho - \frac{\rho +
z}{\sqrt{3}} , 2 \sqrt{3} \rho - \frac{\rho + z}{\sqrt{3}}, \frac{z
- \rho }{\sqrt{3}} \right)
\nonumber\\
&& + e^{\frac{i}{3}(\rho - z)^2} \left( \rho + 2i \right) g^{*}
\left( \sqrt{3} \rho - \frac{\rho - z}{\sqrt{3}} , 2 \sqrt{3} \rho -
\frac{\rho - z}{\sqrt{3}}, \frac{ \rho -z }{\sqrt{3}} \right)
\nonumber\\
&& - \left[ e^{i(\rho + z)^2} \sqrt{\frac{\pi}{2}}\frac{1-i}{2}
\left( erf \left( \frac{1+i}{\sqrt{2}} \left( \rho - z \right)
\right) - erf \left(- \frac{1+i}{\sqrt{2}} z \right) \right) \right.
 \nonumber\\
&& \left. -  e^{-i(\rho + z)^2} \sqrt{\frac{\pi}{2}}\frac{1-i}{2}
\left( erf \left( \frac{1+i}{\sqrt{2}} \left( \rho + z \right)
\right) - erf \left(  \frac{1+i}{\sqrt{2}} z \right) \right) \right]
\nonumber\\
&& + e^{i(\rho + z)^2} \left[ \rho (1 - 2i) - 2 i z \right] g^{*}
\left( 2 \rho + z, 3 \rho + z, -  \rho - z \right)
\nonumber\\
&& + e^{-i(\rho - z)^2} \left[ \sqrt{\frac{\pi}{2}} (1+i) \left( erf
\left( \frac{1+i}{\sqrt{2}} (3 \rho -z) \right) - erf \left(
\frac{1+i}{\sqrt{2}} (2 \rho -z) \right) \right)  \right.
\nonumber\\
&& \left. + 2i (\rho -z) g^{*} \left( 2 \rho - z, 3 \rho - z,
z-\rho
 \right) - \rho g^{*}\left( 2 \rho - z, 3 \rho - z, - z + \rho
 \right) \right]
\nonumber\\
&& + e^{i\frac{z^2}{2}} \frac{\sqrt{\pi}}{2} \frac{1 - i}{2} \left(
erf \left( \frac{1+i}{\sqrt{2}} \left( \sqrt{2} \rho +
\frac{z}{\sqrt{2}} \right) \right) - erf \left( \frac{1+i}{\sqrt{2}}
\left( \sqrt{2} \rho - \frac{z}{\sqrt{2}} \right) \right) \right)
\nonumber\\
&&+\frac{1}{2\rho} \left[ 2 e^{-i\rho(\rho-2z)} -
e^{-4i\rho(2\rho-z)} - 2 e^{-i\rho(3\rho+2z)} + e^{-4i\rho(2\rho+z)}
- 2 e^{-i\rho(\rho+2z)} + e^{-4i\rho (2\rho+z)} \right.
\nonumber\\
&&
\left.\left.
- 2 e^{-i(2z^2+5\rho^2-6\rho z)} + e^{-2i(z^2+5\rho^2-4\rho z)}
\right]
\right\}
\end{eqnarray}
for the other terms.
Finally for the third term in Eq.~\eqref{eq:Pi9ter} the
$q$-integration between $0$ and $k_F$ yields:
\begin{eqnarray}
\label{eq:Pi14}
&& \Pi^0_{(3)}(x,0) \theta (t_x)  = i  \frac{2}{(2
\pi)^4} \frac{\theta (t_x)}{\tau^3} \frac{1}{z}
\Bigg \{ - \frac{\sin (2 \rho z)}{i \rho} + \sqrt{\pi} \frac{1 +i}{2} \Big [ erf \big( \frac{1 - i}{2} (\rho +z) \big)
+ erf (z \rightarrow -z) \Big ]  \nonumber\\
&& \times
\big ( e^{- \frac{i}{2} (\rho - z)^2} - e^{- \frac{i}{2} (\rho + z)^2}    \big ) + \bigg [ - (-z + \rho (1-i)) e^{- \frac{i}{2} (\rho - z)^2} g \bigg (-
\frac{\rho - z}{\sqrt{2}}, \frac{\rho + z}{\sqrt{2}} , \frac{\rho -
z}{\sqrt{2}} \bigg ) + (z \rightarrow - z)  \bigg ]  \nonumber\\
&& + \frac{\sin (2 \rho z)}{\rho} e^{- 4 i \rho^2} + \sqrt{\pi}
\frac{1-i}{2} \bigg [ e^{\frac{i}{2} (\rho - z)^2} \Big ( erf \big(
\frac{1 + i}{2} (3 \rho - z) \big ) - erf \big( \frac{1 + i}{2}
(\rho -z) \big )  \Big ) - (z \rightarrow - z) \bigg ]
\nonumber\\
&&
- \big ( e^{\frac{i}{2} (\rho - z)^2} g^* \bigg ( \frac{\rho - z}{\sqrt{2}}, \frac{3 \rho - z}{\sqrt{2}}, \frac{\rho - z}{\sqrt{2}} \bigg ) - (z \rightarrow - z) \big ) + \frac{1 - \cos (4 \rho z)}{2 z} \Bigg \} ~.
\end{eqnarray}
It is worth noticing that the above formulas embody the physics of diffraction (through
the familiar $erf$) and the attenuation of the propagator (through the Meijer $g$-functions).

\end{document}